\documentclass{JHEP3}
\usepackage{graphics}
\usepackage{graphicx}
\usepackage{amsmath}
\usepackage{cite}
\def\be{\begin{equation}}
\def\ee{\end{equation}}
\def\bea{\begin{eqnarray}}
\def\eea{\end{eqnarray}}
\def\eps{\epsilon}

\def\eps{\epsilon}

\def\cN{  {\cal N}  }
\def\cO{  {\cal O}  }

\preprint{
SI-HEP-2012-12}
\title{Systematics of the cusp anomalous dimension}

\author{Johannes M.\ Henn$^{a}$, Tobias Huber$^{b}$\\
$^a$ Institute for Advanced Study, Princeton, NJ 08540, USA\\
$^b$ Theoretische Physik 1, Naturwissenschaftlich-Technische Fakult\"at,
\\ Universit\"at Siegen, Walter-Flex-Strasse 3, D-57068 Siegen, Germany\\

\email{jmhenn@ias.edu, \\ huber@tp1.physik.uni-siegen.de}}

\abstract{
We study the velocity-dependent cusp anomalous dimension in 
supersymmetric Yang-Mills theory.
In a paper by Correa, Maldacena, Sever, and one of the present authors,
a scaling limit was identified in which the ladder diagrams are dominant
and are mapped onto a Schr\"{o}dinger problem.
We show how to solve the latter in perturbation theory and provide an
algorithm to compute the solution at any loop order. The answer is 
written in terms of harmonic polylogarithms.
Moreover, we give evidence for two curious properties of the result. 
Firstly, we observe that the result can be written using a subset of harmonic polylogarithms
only, at least up to six loops. Secondly, we show that in a light-like
limit, only single zeta values appear in the asymptotic expansion, again
up to six loops.
We then extend the analysis of the scaling limit to systematically include
subleading terms. This leads to a Schr\"{o}dinger-type equation,
but with an inhomogeneous term. 
We show how its solution can be computed in perturbation theory,
in a way similar to the leading order case.
Finally, we analyze the strong coupling limit of these subleading contributions and
compare them to the string theory answer. We find agreement between
the two calculations.
}

\keywords{Supersymmetric gauge theory, NLO Computations}

\begin{document}

\section{Introduction}\label{sec:intro}

The cusp anomalous dimension $\Gamma_{\rm cusp}(\phi)$ was originally
introduced in \cite{PolyakovCusp} as the ultraviolet (UV) divergence of a
Wilson loop with a cusp with Euclidean angle $\phi$. It describes a wide range of 
interesting physical situations. 
It was computed in QCD to the two-loop order in ref. \cite{Korchemsky:1987wg} 
and rederived and simplified in ref. \cite{Kidonakis:2009ev}.

In supersymmetric Yang-Mills theories such as $\cN=4$ super Yang-Mills, 
one can define a Wilson loop operator that couples to scalars in addition to the gauge field \cite{Maldacena:1998im,Rey:1998ik}.
It is natural to consider a loop where the coupling to the scalars is different 
on the two segments of the cusp (but constant along each segment).
The jump in the internal coupling to the scalars is characterized by
an angle $\theta$. The perturbative calculation of this supersymmetric cusp anomalous
dimension $\Gamma_{\rm cusp}(\phi,\theta)$ is similar to the QCD case. It has
been performed to two loops in refs. \cite{Makeenko:2006ds,Drukker:2011za}.
At strong coupling, it is known to second order in the strong coupling expansion \cite{Drukker:2011za}.

Recently, there has been a lot of progress in understanding  $\Gamma_{\rm cusp}(\phi,\theta)$, in various domains.

In a small angle limit, an exact result was found in \cite{Correa:2012at}, based on localization techniques. The exact formula
is in perfect agreement with perturbative results and the result at strong coupling. The same exact formula has
also been obtained in \cite{Fiol:2012sg}.

In ref. \cite{Correa:2012nk}, a relation of the cusp anomalous dimension to the Regge limit of massive scattering amplitudes
was exploited to compute its three-loop value. The relation to scattering amplitudes \cite{Henn:2010bk}, which is valid
in the planar limit, implies in particular that the integrand needed to compute the cusp anomalous
dimension can be deduced from the (in principle known) integrand for planar four-particle scattering amplitudes 
\cite{ArkaniHamed:2010kv,Bourjaily:2011hi,Eden:2012tu}, when appropriately extended to the massive case \cite{Alday:2009zm,Henn:2010bk}.

Very recently, Thermodynamic Bethe Ansatz (TBA) equations have been derived for the cusp anomalous dimensions \cite{IntegrabilityPaper,Drukker:2012de}, and passed
highly non-trivial consistency checks at the three-loop level \cite{IntegrabilityPaper}.

In \cite{Correa:2012nk} a new scaling limit involving the complexified angle $\theta$ was introduced, 
\begin{align}\label{intro-scaling}
i \theta \gg 1, \qquad \lambda \ll 1 \,,\qquad  {\rm with} \qquad \hat{\lambda} = \lambda e^{i \theta}/4 \quad {\rm finite} \,.
\end{align}
Here $\lambda = g^2 N$ is the `t Hooft coupling.
In this limit, the coupling of the loop to the scalars becomes dominant, and the leading order (LO)
contribution is given by simple ladder diagrams, where the rungs of the ladder are scalar
exchanges. It is important to realize that this is a gauge-invariant statement.
The ladder diagrams can be described conveniently using Bethe-Salpeter equations.
The latter are very convenient, since they provide a simple description. They can be solved exactly
in the small angle limit, and it is easy to extract their strong coupling behaviour, finding agreement with the
corresponding string theory calculation.

In this paper, we continue the analysis of the scaling limit of \cite{Correa:2012nk} and initiate a systematic study of
the subleading contributions.
A first question that one faces when computing $\Gamma_{\rm cusp}(\phi, \theta)$ in
perturbation theory is what functions the result can be expressed in.
It is easy to see that the $\theta$ dependence is very simple, and to describe the
$\phi$ dependence the variable $x=e^{i \phi}$ is useful. Experience shows
that in that variable one obtains certain polylogarithms, multiplied by rational prefactors.
In general, it is not known what class of polylogarithms, or more generally what class
of iterated integrals, is sufficient to describe a given problem.

Similar questions are of great current interest in the understanding of the structure
of scattering amplitudes, a problem that is closely related. To phrase the question in that
language, given a  loop integral depending on $n$ space-time points, what is the set of 
functions describing it? 
On the one hand, one could argue that with increasing loop order, integrals
with some number of external points  ``know about''  lower-loop integrals with more 
external points that they contain as subdiagrams (which may e.g.\ contain elliptic integrals), 
making them very complicated,
and perhaps requiring a larger functional basis at higher loops.
On the other hand, one might argue that the set of functions should 
ultimately be determined by the external kinematics of the problem.
An argument in favour of this point of view is that integrals are determined to a great part by their 
singularities, and the location of the latter is intimately tied to the external data. 
These questions are also of enormous practical importance, as they sometimes allow
to make an ansatz for a given problem within a restricted class of functions, see \cite{Dixon:2011pw,CaronHuot:2011kk,Dixon:2012yy}
for recent examples.

In the present case of a single scale problem, it was observed in \cite{Correa:2012nk}
that all functions occurring to the three-loop order could be expressed 
in terms of harmonic polylogarithms, i.e.\ in terms of iterated integrals with
integration kernels $1/x, 1/(1+x), 1/(1-x)$. The fact that this was possible
not only for the final answer, but also for individual loop integrals, and 
in fact also for all integrals of this type found in the literature, seems to
suggest that this is a more general feature. Can this be proven rigorously?
In this paper, we make a first step into this direction. 
We show that this property holds for the LO term of $\Gamma_{\rm cusp}$ in the scaling limit (\ref{intro-scaling}),
and for one of the two contributions at NLO, at any loop order.

We also present an algorithm that determines $\Gamma_{\rm cusp}$ at LO in the scaling limit
at any loop order in terms of harmonic polylogarithms.
As an application, we verified the result of \cite{Correa:2012nk} at three
loops, and evaluated the four-, five-, and six-loop results, which are new.

These results suggest two further properties. First, we find that at least up to six
loops, one can express the result in terms of harmonic polylogarithms
(HPLs)~\cite{Remiddi:1999ew} of argument $x^2$ and indices $0,1$ only.
Second, in the $x \to 0$ limit we find that, again up to six loops, single zeta
values and products thereof are sufficient to describe the coefficients of the asymptotic expansion.

We then discuss NLO terms in the scaling limit.
We show that there are two classes of diagrams that satisfy a slightly
modified Bethe-Salpeter equation.
For one of the two classes of integrals, we show how to construct
the solution in terms of HPLs at any loop order.
For the second class of integrals, we compute the non-trivial integration kernel,
which allows to express the result in terms of iterated integrals having the 
correct degree. We leave the question of whether the latter can be expressed in
terms of HPLs to future work.

We also discuss the strong coupling limit of the Bethe-Salpeter equations,
and compute the scaling limit of the corresponding string theory result.
Under certain assumptions, we find perfect agreement between the two
calculations.

This paper is organized as follows. We begin by reviewing the definition
of the cusp anomalous dimension and the scaling limit in section 2.
Then, in section 3, we present the perturbative solution at leading order in the scaling limit 
to any loop order. We prove that the result can be written in terms of HPLs, and
make further observations about their structure.
In section 4, we discuss the NLO Bethe-Salpeter equations.
In section 5, we take the strong coupling limit of the equations, and compare
them to the corresponding string theory calculation.
There are several appendices containing technical details.


\section{General structure of the Bethe-Salpeter equations at LO and NLO}

Here we discuss the general structure of the Bethe-Salpeter equations at leading order (LO) and next-to-leading order (NLO)
in the scaling limit. The LO equations were already discussed in \cite{Correa:2012nk}. They are a natural generalization of the equations for the quark-antiquark potential \cite{Erickson:1999qv}. 
Here we briefly review the main points. 

We recall the definition of the locally supersymmetric Wilson loop operator in $\cN=4$ super Yang-Mills,
\begin{align}
W \sim {\rm Tr}[ P e^{i \oint A\cdot dx + \oint |dx| \vec{n} \cdot \vec{\Phi} } ]\,,
\end{align}
where $\vec{n}$ is a point on $S^{5}$. The contour we consider consists of two (infinite) segments
forming a cusp of Euclidean angle $\phi$. We take the coupling to the scalars to be constant along
each segment, but with a jump of angle $\theta$ at the cusp, i.e. $\cos \theta = \vec{n}\cdot \vec{n}'$, where
$\vec{n}$ and $\vec{n}'$ are the directions of the two segments.
Such a cusped Wilson loop in general has a logarithmic divergence that takes the form
\begin{align}\label{defgammacusp}
\langle W \rangle \sim e^{- \Gamma_{\rm cusp}(\phi, \theta) \log \frac{\Lambda_{\rm UV}}{\Lambda_{\rm IR}}} \,,
\end{align}
where $\Lambda_{\rm IR/UV}$ are infrared and ultraviolet cutoffs, respectively.
This defines the cusp anomalous dimension $\Gamma_{\rm cusp}(\phi, \theta)$.\footnote{Of course, $\Gamma_{\rm cusp}$ is also a
function of the 't Hooft coupling $g^2 N$, and the number of colours $N$.}

In the scaling limit (\ref{intro-scaling}), the scalar coupling of the loop becomes dominant.
At leading order (LO) in the limit, the segments of the Wilson loop couple to conjugate scalars,
and we need to consider scalar exchange diagrams only. At next-to-leading order (NLO), we
have mostly scalar exchanges, plus one-loop interaction diagrams.

{}
\FIGURE[t]{
\includegraphics[width=1.0\textwidth]{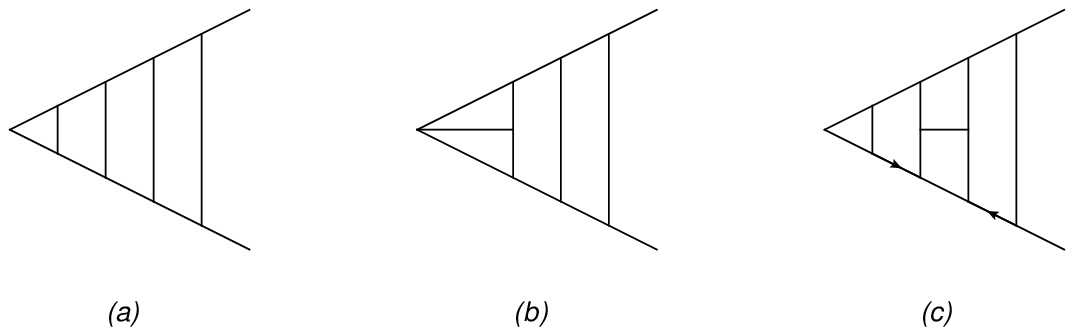}
\caption{Classes of loop integrals contribution to LO (diagram (a)) and NLO (diagram (b) and (c)) in the scaling limit (\ref{intro-scaling}).
Each class can have an arbitrary number of rungs. The arrows in (c) denote a numerator factor $(p+q)^2$, where $p^\mu$ and $q^\mu$ are the momenta along the arrows.}
\label{fig:integrals_large_xi}
}

An analysis of the integrals contributing to the cusp anomalous dimension
allows one to see that the effective diagrams shown in Fig.~\ref{fig:integrals_large_xi} are needed at LO and NLO
in the scaling limit. Since only one-loop internal graphs are allowed at NLO order, one can
deduce the all-loop structure of these corrections already from the known three-loop expression.
The fact that one has effective diagrams that arise after cancellations between various 
gauge-dependent Feynman diagrams\footnote{In ref. \cite{Bykov:2012sc}, this one-loop calculation was 
explicitly performed (for the quark-antiquark potential, corresponding to $\phi \to \pi$), 
in agreement with the result here. Integral class (b) discussed here follows
from a boundary term at the cusp that is absent for the quark-antiquark potential.} 
is intimately related to the similar diagrams appearing in scattering amplitudes.
We illustrate this relation at the level of the loop integrals/integrands in Appendix~C.

{}
\FIGURE[t]{
\includegraphics[width=1.0\textwidth]{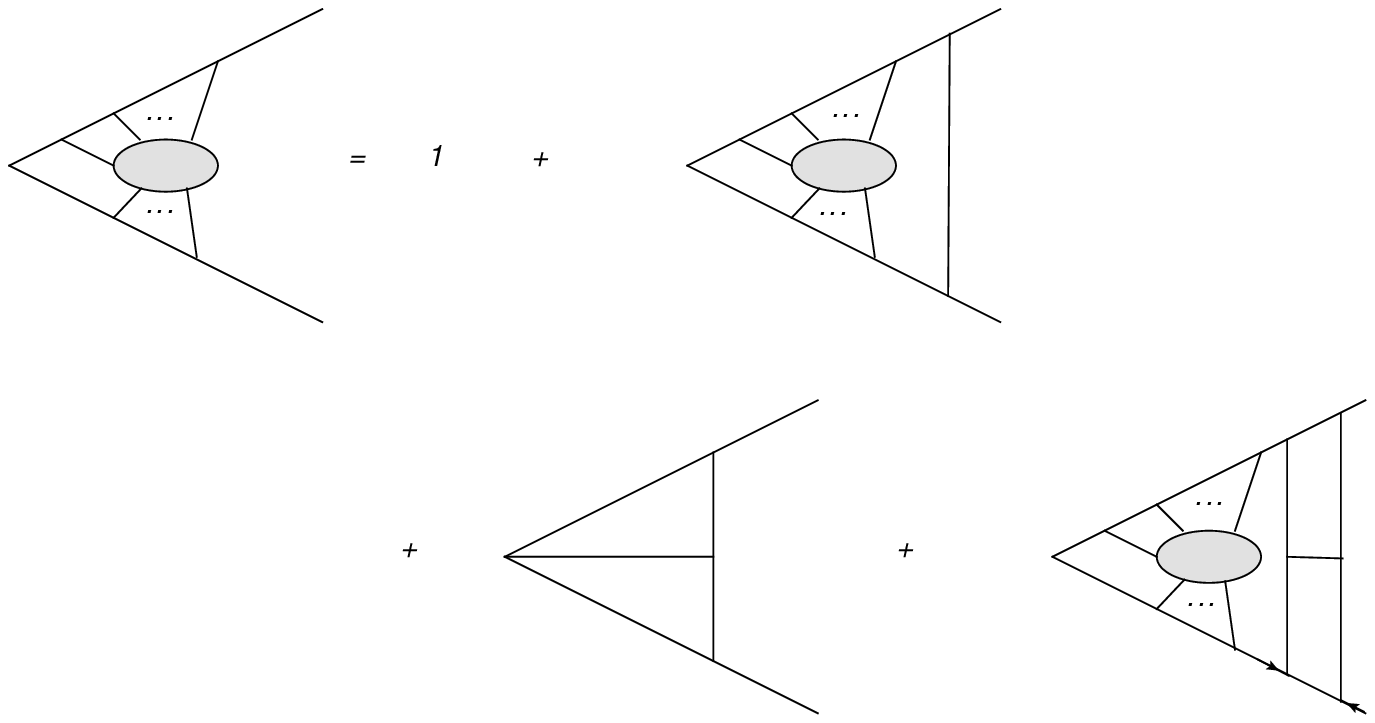}
\caption{Bethe-Salpeter equation at LO and NLO. The arrows denote a numerator factor $(p+q)^2$, with
$p^\mu, q^\mu$ being the momenta flowing along the arrows (in momentum space).}
\label{fig:BetheSalpeter}
}

It is easy to see that the integrals of Fig.~\ref{fig:integrals_large_xi} are described by a Bethe-Salpeter equation.
The latter is shown (schematically) in Fig.~\ref{fig:BetheSalpeter}.
This equation sums the diagrams to all orders in the coupling.
At LO in the scaling limit, only the first line contributes, as the second lines gives contributions of order
$\alpha = \lambda/\hat{\lambda}$ and higher. At NLO, we keep the terms in the second line and compute
the answer linear in $\alpha$. Note that there are also higher-order terms in $\alpha$ contained in this
equation that will only become relevant once we include all NNLO and higher terms.

We can see that there are two new features w.r.t.\ LO. First, the first term of the second line of Fig.~\ref{fig:BetheSalpeter}
is the starting point for the new infinite class of diagrams shown in Fig.~\ref{fig:integrals_large_xi}(b). These terms are absent in
the quark-antiquark potential \cite{Bykov:2012sc}. Second, there is a new interaction term that is a higher-loop
generalization of the simple scalar exchange at LO.

Let us illustrate the usefulness of the Bethe-Salpether equation by reviewing the LO case.
We denote the sum of the ladder diagrams by $F(s,t)$, 
where $-s p^{\mu}$ and $t  q^{\mu}$ are positions on the cusp formed by the momenta $p^{\mu}$
and $q^{\mu}$.
Let us normalize $p^2 = q^2=1$ for convenience.
Note that $F$ also depends on the angle $\phi$ defined by $\cos \phi = p \cdot q$.
Then $F$ satisfies the Bethe-Salpeter equation
\begin{align}
F(S,T) = 1 + \int_0^{S} \, ds \, \int_0^T \,dt \, F(s,t) \, P(s,t) \,,
\end{align}
where 
\begin{align}
P(s,t) = \frac{\hat{\lambda}}{4 \pi^2} \, \frac{1}{ s^2 + t^2 + 2 s t \cos \phi} 
\end{align}
is the propagator corresponding to a scalar exchange.
Changing variables according to $s=e^{\sigma}, t=e^{\tau}$, this becomes
\begin{align}\label{bethe_salpeterLO}
F(\sigma,\tau) = 1 + \int_{-\infty}^{\sigma} \, d\sigma_1 \int_{-\infty}^\tau \, d\tau_1 \, F(\sigma_1,\tau_1) \, P(\sigma_1, \tau_1)\,,
\end{align}
where
\begin{align}
P(\tau,\sigma) = \frac{\hat{\lambda}}{8 \pi^2} \frac{1}{\cosh(\tau-\sigma) + \cos \phi}\,.
\end{align}
Differentiating eq. (\ref{bethe_salpeterLO}), we obtain,
\begin{align}
\partial_{\tau} \partial_{\sigma} F(\sigma,\tau) = F(\sigma,\tau) P(\sigma,\tau) \,.
\end{align}
Let us change variables $y_1 = \tau-\sigma$ and $y_2 = (\tau+\sigma)/2$. 
We can extract $\Gamma_{\rm cusp}$ from the large $y_2 $ behaviour of $F$, 
due to the equivalence of IR and UV divergences, see eq. (\ref{defgammacusp}).

For large $y_{2}$, we can make an ansatz
\begin{align}\label{anstazlargey2}
F = \sum_{n} e^{- \Omega_{n} y_{2}} \Psi_{n}(y_{1}) \,.
\end{align}
We are interested in the leading term, corresponding to the lowest eigen-energy $\Omega_{0}$.
Using the ansatz (\ref{anstazlargey2}), one finds \cite{Correa:2012nk}
\begin{align}\label{schroedingerLO}
& \left[ -\partial_{y_1}^2 - \frac{\hat{\lambda}}{8 \pi^2} \frac{1}{(\cosh y_1 + \cos \phi)} + \frac{\Omega^2(\phi)}{4} \right] \Psi(y_1 ,\phi) 
= 0 \,.
\end{align}
This is a one-dimensional Schr\"{o}dinger  problem. The ground state energy $\Omega_{0}$
is related to the cusp anomalous dimension in the scaling limit through $\Gamma_{\rm cusp} = - \Omega_{0}$.

In summary, the Bethe-Salpeter equation has allowed us to conveniently sum an infinite class of
diagrams. As a result, extracting the remaining overall logarithmic divergence could be done in a simple
way, and the remaining calculation does not require any regulator. Moreover, the structure of the equation
allowed us to rewrite the problem in terms of a linear differential equation.

We will now solve this equation in perturbation theory.
In section \ref{sec:subleading}, we will discuss the effects of the two new features that appear at NLO.

\section{Solution to the scaling limit at leading order}
\label{sec:LO}

\subsection{Setup}
To obtain the perturbative solution of (\ref{schroedingerLO}), we follow \cite{Correa:2012nk} and perform the 
change of variables
\begin{align}
\Psi(y_1) = \eta(y_1) e^{-\Omega_0 y_1 /2}
\end{align}
The exponential factor gives the correct solution as $y_1 \to \infty$, and we can normalize $\eta(y_{1} = \infty)=1$.
We can determine $\Omega_{0}$ from $\eta$ thanks to the boundary condition
\begin{align}\label{boundary0}
\partial_{y_1} \Psi(y_1 ) |_{y_1 = 0} = 0\,,
\end{align}
which follows from the $y_1 \to -y_1$ symmetry of the problem.
Defining a new variable $w = e^{-y_1}$, and $x=e^{i \phi}$, 
the boundary condition (\ref{boundary0}) becomes 
\begin{align}\label{relation_omega_eta}
\Omega_{0}(x) = -2 w \partial_w \log \eta(w,x) {|}_{w=1}\,,
\end{align}
and the  Schr\"{o}dinger  equation (\ref{schroedingerLO}) reads
\begin{align}\label{schroedinger}
\partial_w w \partial_w \eta = -\Omega_0(x) \partial_w \eta + \hat\kappa \left[ \frac{1}{w+x^{-1}} - \frac{1}{w+x}  \right] \eta \,,\qquad \hat\kappa = \frac{\hat\lambda \, x }{4 \pi^2 (1-x^2) }
\end{align}
The wavefunction $\eta$ can be obtained by integrating the Schr\"{o}dinger equation iteratively in the coupling,
$\Omega_0 = \hat\kappa \Omega_0^{(1)} + \hat\kappa^2 \Omega_0^{(2)} + \ldots$, and $\eta = 1+ \hat\kappa \eta^{(1)} + \ldots$.
Let us now analyze in detail the perturbative solution for $\eta$ and $\Omega$.

\subsection{Iterative solution}
It is convenient to introduce an abbreviation for the nested integrals that one encounters in this problem.
In analogy to two-dimensional harmonic polylogarithms (2dHPLs),
we are going to use the self-explanatory notation
\begin{align}
H_{V}(w,x)=& \int_0^1 dw'  \left[ \frac{1}{w' + 1/(wx)} - \frac{1}{w'+x/w} \right] \,,
\end{align}
and 
\begin{align}
H_{V, \vec{b}}(w,x) =& \int_0^1 dw'  \left[ \frac{1}{w' + 1/(wx)} - \frac{1}{w'+x/w} \right]  H_{\vec{b}}(w' w,x) \,, \label{defHV0}\\
H_{0, \vec{b}}(w,x) =&  \int_0^1 \frac{dw'}{w'} \,  H_{\vec{b}}(w' w,x) \,.
\end{align}
In the following we will sometimes drop the arguments $(w,x)$ for brevity.
So in general we will have $H_{\vec{b}}$, where the weight vector $\vec{b}$ has entries $V$ and $0$, with $0$ not appearing in the last entry.

It is straightforward to write the perturbative answer for $\eta$ in terms of these integrals. 
We find
\begin{align}
\eta^{(1)} =& H_{0,V} \\
\eta^{(2)} =& H_{0,V,0,V}-H_{0,0,V} \Omega_{0}^{(1)} \,,
\end{align}
and so on.
Using eq. (\ref{relation_omega_eta}) we find
\begin{align}
\Omega^{(1)}_{0} =& - 2 H_{V} \,, \label{om1}\\
\Omega^{(2)}_{0} =& -2 H_{V} H_{0,V} -2 H_{V,0,V} \,, \label{om2}\\
\Omega^{(3)}_{0} =& -2 H_{V} H_{0, V}^2 - 4 H_{V}^2 H_{0, 0, V}  
- 2 H_{0, V} H_{V, 0, V}  \nonumber \\
&- 2 H_{V} H_{0, V, 0, V} - 
 4 H_{V} H_{V, 0, 0, V} - 2 H_{V, 0, V, 0, V}\,,\label{om3}
\end{align}
etc. These last relations are understood at $w=1$.

In principle, eqs. (\ref{om1}), (\ref{om2}), (\ref{om3}), and their higher-order analogues,
together with $\Gamma = - \Omega_{0}$, provide formulas for $\Gamma$. 
However, this representation is clearly not an optimal one.
In the following, we will simplify it by converting it to a more appropriate and
simpler class of iterated integrals. This will also allow us to make further
observations regarding the structure of the result.

\subsection{Structure of the perturbative result}
Here, we first show certain properties of $\eta$ and $\Omega_0$, 
and then outline an algorithm for expressing $\Omega_0$ in terms of harmonic 
polylogarithms.

As we show presently, the total differential of $\eta$ at any loop order is 
of the form
\begin{align}\label{structure_eta}
d \eta^{(L)} =& f_1 \, d \log x + f_2 \, d \log (1+x) + f_3 \, d \log (1-x)   \nonumber \\
& + f_4 \, d \log (w+x) + f_5 \, d \log (w+1/x) \,,
\end{align}
with the $f_{i}$ being functions of the same type as $\eta^{(L)}$, but
of degree (i.e. number of iterated integrals) lowered by one.
From equations (\ref{structure_eta}) and (\ref{relation_omega_eta}) it immediately follows that
\begin{align}\label{structure_omega}
d \Omega^{(L)} = g_1 \, d \log x + g_2 \, d \log (1+x) + g_3 \, d \log (1-x)  \,,
\end{align}
with $g_{i}$ being functions of degree lowered by one, and satisfying the same property. 
This,  implies that at any loop order $L$, $\Omega^{(L)}$ can be expressed in terms of harmonic polylogarithms (HPLs)
of degree $(2 L-1)$.

The latter are defined iteratively by
\begin{align}\label{defHPL}
H_{a_1 , a_2 , \ldots, a_n }(x) = \int_0^x f_{a_1}(t) H_{a_2, \ldots, a_n }(t) \, dt \,,
\end{align}
where the integration kernels are
\begin{align}
f_{1}(x) = \frac{1}{1-x} \,,\qquad f_{0}(x) = \frac{1}{x} \,,\qquad f_{-1}(x) = \frac{1}{1+x} \,.
\end{align}
The degree-one functions needed to start the recursion are defined as
\begin{align}
H_{1}(x) = - \log(1-x) \,,\qquad H_{0}(x) = \log(x) \,,\qquad H_{-1}(x) = \log(1+x) \,.
\end{align}
The subscript of $H$ is called the weight vector. A common abbreviation is to replace occurrences
of $m$ zeros to the left of $\pm1$ by $\pm(m+1)$. For example, $H_{0,0,1,0,-1}(x) = H_{3,-2}(x)$.

Note that a corollary of equations (\ref{structure_eta}) and (\ref{structure_omega}) is that
the symbol \cite{symbols1,symbols2} of $\eta$ is constructed from a five-letter alphabet consisting of $x, 1\pm x, w+x, w+1/x$.
Similarly, the symbol of $\Omega_{0}$ is constructed from the three-letter alphabet $x,1\pm x$.
Of course, knowing the full differential provides us with much more information than just the symbol.

In order to prove the above statements, let us point out a relation of the $H_{\vec{b}}(w,x)$ to a known, 
albeit more general class of functions, the Goncharov polylogarithms~\cite{Goncharov},
\begin{align}
G(a_1,\ldots a_n ; z) = \int_0^z \frac{dt}{t-a_{1}} G(a_{2}, \ldots ,a_{n}; t) \,,
\end{align}
with 
\begin{align}
G(a_1 ;z) = \int_0^z \frac{dt}{t-a_{1}}  \,.
\end{align}
In our case, the $a_{i}$ are taken from $\{0, -x, -1/x\}$ and $z=w$.
For example, we have
\begin{align}
H_{V}(w,x) =& G(-1/x;w) - G(-x;w) \,,\\
H_{0,V}(w,x) =& G(0,-1/x;w) - G(0,-x;w) \,, \label{exampleH0V}\\
H_{V,0,V}(w,x) =& G(-1/x,0,-1/x;w) -G(-1/x,0,-x;w)\nonumber \\
&+G(-x,0,-x;w)-G(-x,0,-1/x;w) \,,
\end{align}
and so on.
The total differential of a general Goncharov polylogarithm is 
\begin{align}\label{differentialG}
d G(a_1 , \ldots a_{n};z) =&\; G(\hat{a}_{1}, a_{2}, \ldots a_{n}; z) \, d \log\frac{z-a_1}{a_1 - a_2} \nonumber \\
& + G(a_{1}, \hat{a}_{2}, a_{3} , \ldots, a_{n}; z) \,  d \log \frac{a_1 - a_2 }{a_2 - a_3} + \ldots +  \nonumber \\
& + G(a_{1}, \ldots, a_{n-1}, \hat{a}_{n}; z) \, d \log \frac{a_{n-1}-a_{n}}{a_{n}} \,,
\end{align}
where $\hat{a}$ means that this element is omitted.

Given the possible values of the $a_{i}$ in our case, it is straightforward to verify eq. (\ref{structure_eta}).

\subsection{Rewriting the expressions for $\Omega_{0}$ in terms of HPLs}

We have proven that $\Omega_{0}$ can be written in terms of HPLs. 
Let us now explain how to find explicit results in terms of HPLs.
We will begin by a simple example, and then outline an algorithm for doing so in general.

We observed that eq.~(\ref{differentialG}), when applied to any function $H_{\vec{b}}(w=1,x)$
gives a result of the form (\ref{structure_omega}). Iterating this procedure
for the lower degree functions $g_{i}$ in that equation, together with the fact that
at any order we have a boundary condition at $w=1$, gives us the complete
information for that function, in a form that makes contact with the definition of HPLs,
see eq. (\ref{defHPL}).

As an example, let us write $H_{0,V}(1,x)$ in terms of HPLs.
According to eq. (\ref{exampleH0V}), we need to rewrite $G(0,-x;1)$ and
$G(0,-1/x;1)$ in terms of HPLs.
Specializing (\ref{differentialG}) to the present case we have
\begin{align}
d \,G(0,-x;1) =& \, - G(-x;1) \, d \log x  \nonumber \\
 =& \, - \log ((1+x)/x) \, d \log x \nonumber \\
 =& \,  - \left[ H_{-1}(x) - H_{0}(x) \right] \, d\log x \,.
\end{align}
The integration can be done using the definition (\ref{defHPL}),
\begin{align}
G(0,-x;1) =& -H_{0,-1}(x) + H_{0,0}(x) + C \,.
\end{align}
It is convenient to relate $C$ to the value at $x=1$,
\begin{align}\label{exampleHPL1}
G(0,-x;1) =&  -H_{0,-1}(x) + H_{0,0}(x) + \frac{1}{2} \zeta_2 + G(0,-1;1) \,,
\end{align}
where we used that $H_{0,-1}(1) = 1/2 \zeta_2$.

Similarly, we find
\begin{align}\label{exampleHPL2}
G(0,-1/x;1) =& H_{0,-1}(x) -\frac{1}{2} \zeta_2 + G(0,-1;1)\,.
\end{align}
Combining eqs. (\ref{exampleHPL1}) and (\ref{exampleHPL2}) we find
\begin{align}
H_{0,V}(1,x) = 2 H_{0,-1}(x) - H_{0,0}(x) - \zeta_{2}\,.
\end{align}
By construction, $H_{0,V}(1,1)=0$.

In summary, from this example it becomes clear how to rewrite any of the functions
occurring in our problem in terms of HPLs, using the following steps:
{\allowdisplaybreaks
\begin{enumerate}
\item Express the functions in terms of Goncharov polylogarithms
\item Use eq. (\ref{differentialG}) in order to compute their (total) differential; 
since all other variables are constants, this gives the derivative in $x$.
\item By iteration, that differential is of the form (\ref{structure_omega}), with the $g_{i}$ appearing there being 
HPLs. It can therefore be integrated in terms of HPLs, using (\ref{defHPL}).
\item Integrate the equation with the boundary term  at $x=1$.
\item Add up all terms; the boundary Goncharov polylogarithms at $x=1$ do not necessarily drop out,
but they are simple since they correspond to harmonic polylogarithms evaluated at $x=1$.
\end{enumerate}
}

Using this algorithm, we find e.g.\ the following expressions that are required to the
three-loop order, 
{\allowdisplaybreaks
\begin{align}
H_{V}(1,x) =& H_{0}(x)\,,\\
H_{V,0,V}(1,x) =& -4 H_{-3}(x) - \zeta_2 H_{0}( x) + 2 H_{-2, 0}( x) - 
 4 H_{2, 0}( x) - H_{0, 0, 0}( x) - 2 \zeta_3\,, \\
 H_{0,0,0,V}(1,x)=& - \frac{7 \pi^4}{360} + 2 H_{-4}( x) - \zeta_2 H_{0, 0}(x) - 
 H_{0, 0, 0, 0}(x)\,, \\
 H_{0,V,0,V}(1,x) =&  \frac{19 \pi^4}{360} - 14 H_{-4}( x) - \pi^2 H_{-2}( x) + 
 \frac{2}{3} \pi^2 H_{2}( x) + 8 H_{-3, -1}( x)  - 4 H_{-3, 0}( x) \nonumber \\ & + 
 12 H_{-2, -2}( x) + \frac{1}{6} \pi^2 H_{0, 0}( x) - 
 8 H_{2, -2}( x) - 6 H_{-2, 0, 0}( x)  \nonumber \\ & + 4 H_{2, 0, 0}( x) + 
 H_{0, 0, 0, 0}( x) + 2 H_{0}( x) \zeta_3 \,, \\
 H_{V, 0, V, 0, V}(1,x) =& 40 H_{-5}(x) - \frac{2 \pi^2}{3} H_{-3}( x) + 
   \frac{19 \pi^4}{360} H_{0}(x) +  \frac{4 \pi^2}{3} H_{3}(x) + 24 H_{-4, -1}( x) \nonumber \\
   &- 38 H_{-4, 0}(x) - 16 H_{-3, -2}( x) - 24 H_{-2, -3}( x) -  \pi^2 H_{-2, 0}( x) 
      + 32 H_{2, -3}( x) \nonumber \\
      &+  \frac{4 \pi^2}{3} H_{2, 0}( x) + 
   32 H_{3, -2}( x) + 52 H_{4, 0} ( x) + 8 H_{-3, -1, 0}( x) - 
   4 H_{-3, 0, 0}( x) \nonumber \\
   &- 16 H_{-3, 1, 0}( x) + 12 H_{-2, -2, 0}( x) - 
   24 H_{-2, 2, 0}( x) + \frac{\pi^2}{6} H_{0, 0, 0}( x) \nonumber \\
   &- 
   16 H_{2, -2, 0}( x)  + 32 H_{2, 2, 0}( x) - 16 H_{3, -1, 0}( x) + 
   8 H_{3, 0, 0}( x) + 32 H_{3, 1, 0}( x)\nonumber \\
   & - 6 H_{-2, 0, 0, 0} ( x) + 
   8 H_{2, 0, 0, 0}( x) + H_{0, 0, 0, 0, 0}( x) +  \frac{\pi^2 \zeta_3 }{3} - 
   12 \zeta_3 H_{-2}(x) \nonumber \\
   &+ 16 \zeta_3 H_{2}( x) + 
   2 \zeta_3 H_{0, 0}( x) + 6 \zeta_5 \,.
\end{align}
}
Plugging these formulas into eq. (\ref{om3}), we find perfect agreement with
the three-loop result of ref. \cite{Correa:2012nk}.

In the next section, we show explicit new results that we obtained using this algorithm.

\subsection{Explicit new results, and further surprises}

Using the method described in the previous section, we explicitly determined $\Omega_{0}^{(1)}(x)$ -- 
$\Omega_{0}^{(6)}(x)$ in terms of HPLs. We will show these formulas below.

When analyzing the resulting formulas, in fact we found a further simplification, that was already noticed
in~\cite{Correa:2012nk} up to the three-loop level. Although results for individual integrals contain in general HPLs with all possible
indices $0,\pm 1$, we observe that, at least up to six loops, it is possible to write the final result
in terms of HPLs having indices $0,1$ only, provided that we use $x^2$ as argument instead of $x$.
That property is manifest in the following formulas. 
Up to three loops, one finds
{\allowdisplaybreaks
\begin{align}
\Omega_{0}^{(1)}(x) =&\;  - H_0 \; , \label{omega1atLO} \\
\Omega_{0}^{(2)}(x) =&\;  4 \, \zeta_{3} + 2 \, \zeta_{2} \, H_{0} + 2 \, H_{2, 0} + H_{0, 0, 0} \; , \\
\Omega_{0}^{(3)}(x) =&\; -8 \, \zeta_{2} \, \zeta_{3} - 12 \, \zeta_{5} - 12 \, \zeta_{4} \, H_{0}
                      - 16 \, \zeta_{3} \, H_{2} - 8 \, \zeta_{2} \, H_{3} - 4 \, \zeta_{3} \, H_{0, 0}
		      - 8 \, \zeta_{2} \, H_{2, 0} \nonumber \\ 
		     &  - 8 \, H_{4, 0} - 8 \, \zeta_{2} \, H_{0, 0, 0}
		      - 8 \, H_{2, 2, 0} - 4 \, H_{3, 0, 0} - 8 \, H_{3, 1, 0} - 4 \, H_{2, 0, 0, 0} - 6 \, H_{0, 0, 0, 0, 0} \; .
\end{align}}
Our result at four loops reads
{\allowdisplaybreaks
\begin{align}\label{omega4atLO}
\Omega_{0}^{(4)}(x) =&\;  48 \, \zeta_{3} \, \zeta_{4} + 24 \, \zeta_{2} \, \zeta_{5} + 36 \, \zeta_{7} + 
 8 \, \zeta_{3}^2 \, H_{0} + 51 \, \zeta_{6} \, H_{0} + 
 48 \, \zeta_{2} \, \zeta_{3} \, H_{2} + 72 \, \zeta_{5} \, H_{2} \nonumber \\ &+ 
 96 \, \zeta_{4} \, H_{3} + 88 \, \zeta_{3} \, H_{4} + 
 80 \, \zeta_{2} \, H_{5} + 32 \, \zeta_{2} \, \zeta_{3} \, H_{0, 0} + 
 20 \, \zeta_{5} \, H_{0, 0} + 72 \, \zeta_{4} \, H_{2, 0}  \nonumber \\ & + 
 96 \, \zeta_{3} \, H_{2, 2} + 48 \, \zeta_{2} \, H_{2, 3} + 
 32 \, \zeta_{3} \, H_{3, 0} + 128 \, \zeta_{3} \, H_{3, 1} + 
 64 \, \zeta_{2} \, H_{3, 2} + 80 \, \zeta_{2} \, H_{4, 0}  \nonumber \\ &+ 
 48 \, \zeta_{2} \, H_{4, 1} + 92 \, H_{6, 0} + 
 114 \, \zeta_{4} \, H_{0, 0, 0} + 24 \, \zeta_{3} \, H_{2, 0, 0} + 
 48 \, \zeta_{2} \, H_{2, 2, 0} + 48 \, H_{2, 4, 0}  \nonumber \\ &+ 
 64 \, \zeta_{2} \, H_{3, 0, 0} + 64 \, \zeta_{2} \, H_{3, 1, 0} + 
 64 \, H_{3, 3, 0} + 80 \, H_{4, 2, 0} + 80 \, H_{5, 0, 0} + 
 80 \, H_{5, 1, 0}  \nonumber \\ &+ 24 \, \zeta_{3} \, H_{0, 0, 0, 0} + 
 48 \, \zeta_{2} \, H_{2, 0, 0, 0} + 48 \, H_{2, 2, 2, 0} + 
 24 \, H_{2, 3, 0, 0} + 48 \, H_{2, 3, 1, 0} + 
 64 \, H_{3, 1, 2, 0} \nonumber \\ & + 32 \, H_{3, 2, 0, 0} + 
 64 \, H_{3, 2, 1, 0} + 64 \, H_{4, 0, 0, 0} + 
 24 \, H_{4, 1, 0, 0} + 48 \, H_{4, 1, 1, 0} + 
 92 \, \zeta_{2} \, H_{0, 0, 0, 0, 0} \nonumber \\ & + 24 \, H_{2, 2, 0, 0, 0} + 
 48 \, H_{3, 0, 0, 0, 0} + 32 \, H_{3, 1, 0, 0, 0} + 
 36 \, H_{2, 0, 0, 0, 0, 0} + 92 \, H_{0, 0, 0, 0, 0, 0, 0} \; .
\end{align}
At five loops we obtain
\begin{align}\label{omega5atLO}
\Omega_{0}^{(5)}(x) =&\; -\frac{32}{3} \, \zeta_{3}^3 - 144 \, \zeta_{4} \, \zeta_{5} - 204 \, \zeta_{3} \, \zeta_{6} - 
 72 \, \zeta_{2} \, \zeta_{7} - \frac{340}{3} \, \zeta_{9} - 64 \, \zeta_{2} \, \zeta_{3}^2 \, H_{0} - 
 80 \, \zeta_{3} \, \zeta_{5} \, H_{0}  \nonumber \\
   &
- \frac{620}{3} \, \zeta_{8} \, H_{0} - 
 384 \, \zeta_{3} \, \zeta_{4} \, H_{2} - 192 \, \zeta_{2} \, \zeta_{5} \, H_{2} - 
 288 \, \zeta_{7} \, H_{2} - 96 \, \zeta_{3}^2 \, H_{3} - 
 612 \, \zeta_{6} \, H_{3} \nonumber \\
   &
- 576 \, \zeta_{2} \, \zeta_{3} \, H_{4} - 
 528 \, \zeta_{5} \, H_{4} - 1776 \, \zeta_{4} \, H_{5} - 
 1216 \, \zeta_{3} \, H_{6} - 1568 \, \zeta_{2} \, H_{7} - 
 456 \, \zeta_{3} \, \zeta_{4} \, H_{0, 0} \nonumber \\
   &
- 144 \, \zeta_{2} \, \zeta_{5} \, 
  H_{0, 0} - 84 \, \zeta_{7} \, H_{0, 0} - 
 64 \, \zeta_{3}^2 \, H_{2, 0} - 408 \, \zeta_{6} \, H_{2, 0} - 
 384 \, \zeta_{2} \, \zeta_{3} \, H_{2, 2} - 576 \, \zeta_{5} \, H_{2, 2} \nonumber \\
   &
- 
 768 \, \zeta_{4} \, H_{2, 3} - 704 \, \zeta_{3} \, H_{2, 4} - 
 640 \, \zeta_{2} \, H_{2, 5} - 384 \, \zeta_{2} \, \zeta_{3} \, H_{3, 0} - 
 240 \, \zeta_{5} \, H_{3, 0} - 576 \, \zeta_{2} \, \zeta_{3} \, H_{3, 1} \nonumber \\
   &
- 
 864 \, \zeta_{5} \, H_{3, 1} - 1152 \, \zeta_{4} \, H_{3, 2} - 
 1056 \, \zeta_{3} \, H_{3, 3} - 960 \, \zeta_{2} \, H_{3, 4} - 
 1656 \, \zeta_{4} \, H_{4, 0} - 1152 \, \zeta_{4} \, H_{4, 1} \nonumber \\
   &
- 
 1440 \, \zeta_{3} \, H_{4, 2} - 1152 \, \zeta_{2} \, H_{4, 3} - 
 704 \, \zeta_{3} \, H_{5, 0} - 1856 \, \zeta_{3} \, H_{5, 1} - 
 1216 \, \zeta_{2} \, H_{5, 2} - 1808 \, \zeta_{2} \, H_{6, 0} \nonumber \\
   &
- 
 960 \, \zeta_{2} \, H_{6, 1} - 2144 \, H_{8, 0} - 
 48 \, \zeta_{3}^2 \, H_{0, 0, 0} - 948 \, \zeta_{6} \, H_{0, 0, 0} - 
 256 \, \zeta_{2} \, \zeta_{3} \, H_{2, 0, 0} - 160 \, \zeta_{5} \, H_{2, 0, 0} \nonumber \\
   &
- 
 576 \, \zeta_{4} \, H_{2, 2, 0} - 768 \, \zeta_{3} \, H_{2, 2, 2} - 
 384 \, \zeta_{2} \, H_{2, 2, 3} - 256 \, \zeta_{3} \, H_{2, 3, 0} - 
 1024 \, \zeta_{3} \, H_{2, 3, 1}\nonumber \\
   &
 - 512 \, \zeta_{2} \, H_{2, 3, 2} - 
 640 \, \zeta_{2} \, H_{2, 4, 0} - 384 \, \zeta_{2} \, H_{2, 4, 1} - 
 736 \, H_{2, 6, 0} - 1368 \, \zeta_{4} \, H_{3, 0, 0}  \nonumber \\
   &
 - 864 \, \zeta_{4} \, H_{3, 1, 0} - 1152 \, \zeta_{3} \, H_{3, 1, 2} - 
 576 \, \zeta_{2} \, H_{3, 1, 3} - 384 \, \zeta_{3} \, H_{3, 2, 0} - 
 1536 \, \zeta_{3} \, H_{3, 2, 1} \nonumber \\
   &
- 768 \, \zeta_{2} \, H_{3, 2, 2} - 
 960 \, \zeta_{2} \, H_{3, 3, 0} - 576 \, \zeta_{2} \, H_{3, 3, 1} - 
 1104 \, H_{3, 5, 0} - 448 \, \zeta_{3} \, H_{4, 0, 0}\nonumber \\
   &
 - 
 384 \, \zeta_{3} \, H_{4, 1, 0} - 1536 \, \zeta_{3} \, H_{4, 1, 1} - 
 768 \, \zeta_{2} \, H_{4, 1, 2} - 1152 \, \zeta_{2} \, H_{4, 2, 0} - 
 576 \, \zeta_{2} \, H_{4, 2, 1} \nonumber \\
   &
- 1392 \, H_{4, 4, 0} - 
 1648 \, \zeta_{2} \, H_{5, 0, 0} - 1216 \, \zeta_{2} \, H_{5, 1, 0} - 
 384 \, \zeta_{2} \, H_{5, 1, 1} - 1648 \, H_{5, 3, 0}\nonumber \\
   &
 - 
 1808 \, H_{6, 2, 0} - 2352 \, H_{7, 0, 0} - 
 1568 \, H_{7, 1, 0} - 368 \, \zeta_{2} \, \zeta_{3} \, H_{0, 0, 0, 0} - 
 152 \, \zeta_{5} \, H_{0, 0, 0, 0}\nonumber \\
   &
 - 912 \, \zeta_{4} \, H_{2, 0, 0, 0} - 
 192 \, \zeta_{3} \, H_{2, 2, 0, 0} - 384 \, \zeta_{2} \, H_{2, 2, 2, 0} - 
 384 \, H_{2, 2, 4, 0} - 512 \, \zeta_{2} \, H_{2, 3, 0, 0} \nonumber \\
   &
- 
 512 \, \zeta_{2} \, H_{2, 3, 1, 0} - 512 \, H_{2, 3, 3, 0} - 
 640 \, H_{2, 4, 2, 0} - 640 \, H_{2, 5, 0, 0} - 
 640 \, H_{2, 5, 1, 0} \nonumber \\
   &
- 288 \, \zeta_{3} \, H_{3, 0, 0, 0} - 
 288 \, \zeta_{3} \, H_{3, 1, 0, 0} - 576 \, \zeta_{2} \, H_{3, 1, 2, 0} - 
 576 \, H_{3, 1, 4, 0} - 768 \, \zeta_{2} \, H_{3, 2, 0, 0} \nonumber \\
   &
- 
 768 \, \zeta_{2} \, H_{3, 2, 1, 0} - 768 \, H_{3, 2, 3, 0} - 
 960 \, H_{3, 3, 2, 0} - 960 \, H_{3, 4, 0, 0} - 
 960 \, H_{3, 4, 1, 0} \nonumber \\
   &
- 1392 \, \zeta_{2} \, H_{4, 0, 0, 0} - 
 768 \, \zeta_{2} \, H_{4, 1, 0, 0} - 768 \, \zeta_{2} \, H_{4, 1, 1, 0} - 
 768 \, H_{4, 1, 3, 0} - 1152 \, H_{4, 2, 2, 0}\nonumber \\
   &
 - 
 1184 \, H_{4, 3, 0, 0} - 1152 \, H_{4, 3, 1, 0} - 
 1216 \, H_{5, 1, 2, 0} - 1376 \, H_{5, 2, 0, 0} - 
 1216 \, H_{5, 2, 1, 0}\nonumber \\
   &
 - 2080 \, H_{6, 0, 0, 0} - 
 1440 \, H_{6, 1, 0, 0} - 960 \, H_{6, 1, 1, 0} - 
 2172 \, \zeta_{4} \, H_{0, 0, 0, 0, 0} - 
 192 \, \zeta_{3} \, H_{2, 0, 0, 0, 0} \nonumber \\
   &
- 
 384 \, \zeta_{2} \, H_{2, 2, 0, 0, 0} - 384 \, H_{2, 2, 2, 2, 0} - 
 192 \, H_{2, 2, 3, 0, 0} - 384 \, H_{2, 2, 3, 1, 0} - 
 512 \, H_{2, 3, 1, 2, 0} \nonumber \\
   &
- 256 \, H_{2, 3, 2, 0, 0} - 
 512 \, H_{2, 3, 2, 1, 0} - 512 \, H_{2, 4, 0, 0, 0} - 
 192 \, H_{2, 4, 1, 0, 0} - 384 \, H_{2, 4, 1, 1, 0} \nonumber \\
   &
- 
 1104 \, \zeta_{2} \, H_{3, 0, 0, 0, 0} - 
 576 \, \zeta_{2} \, H_{3, 1, 0, 0, 0} - 576 \, H_{3, 1, 2, 2, 0} - 
 288 \, H_{3, 1, 3, 0, 0} - 576 \, H_{3, 1, 3, 1, 0}\nonumber \\
   &
 - 
 768 \, H_{3, 2, 1, 2, 0} - 384 \, H_{3, 2, 2, 0, 0} - 
 768 \, H_{3, 2, 2, 1, 0} - 768 \, H_{3, 3, 0, 0, 0} - 
 288 \, H_{3, 3, 1, 0, 0} \nonumber \\
   &
- 576 \, H_{3, 3, 1, 1, 0} - 
 768 \, H_{4, 1, 1, 2, 0} - 384 \, H_{4, 1, 2, 0, 0} - 
 768 \, H_{4, 1, 2, 1, 0} - 960 \, H_{4, 2, 0, 0, 0} \nonumber \\
   &
- 
 288 \, H_{4, 2, 1, 0, 0} - 576 \, H_{4, 2, 1, 1, 0} - 
 1728 \, H_{5, 0, 0, 0, 0} - 1136 \, H_{5, 1, 0, 0, 0} - 
 192 \, H_{5, 1, 1, 0, 0} \nonumber \\
   &
- 384 \, H_{5, 1, 1, 1, 0} - 
 368 \, \zeta_{3} \, H_{0, 0, 0, 0, 0, 0} - 
 736 \, \zeta_{2} \, H_{2, 0, 0, 0, 0, 0} - 
 192 \, H_{2, 2, 2, 0, 0, 0} - 384 \, H_{2, 3, 0, 0, 0, 0} \nonumber \\
   &
- 
 256 \, H_{2, 3, 1, 0, 0, 0} - 288 \, H_{3, 1, 2, 0, 0, 0} - 
 576 \, H_{3, 2, 0, 0, 0, 0} - 384 \, H_{3, 2, 1, 0, 0, 0} - 
 1408 \, H_{4, 0, 0, 0, 0, 0} \nonumber \\
   &
- 576 \, H_{4, 1, 0, 0, 0, 0} - 
 384 \, H_{4, 1, 1, 0, 0, 0} - 
 2144 \, \zeta_{2} \, H_{0, 0, 0, 0, 0, 0, 0} - 
 288 \, H_{2, 2, 0, 0, 0, 0, 0} \nonumber \\
   &
- 1104 \, H_{3, 0, 0, 0, 0, 0, 0} - 
 432 \, H_{3, 1, 0, 0, 0, 0, 0} - 
 736 \, H_{2, 0, 0, 0, 0, 0, 0, 0} - 
 2680 \, H_{0, 0, 0, 0, 0, 0, 0, 0, 0} \; .
\end{align}
}
The six-loop result fills several pages and is therefore relegated to Appendix~D.
All HPLs are understood to have argument $x^2$.
Note that all indices are positive, in other words only the basic indices $0$ and $1$ appear.
This is remarkable, and such a rewriting is in general not possible for individual terms 
contributing to (\ref{omega1atLO})~--~(\ref{omega5atLO}) and~(\ref{omega6atLO}).

It is very remarkable that within each of the equations (\ref{omega1atLO})~--~(\ref{omega5atLO}) and~(\ref{omega6atLO}) all terms
have the same sign, and the common sign is alternating as the loop order increases.
In fact, there is a sign constraint from the fact that the loop integrals
leading to $\Omega_{0}$ should be positive, at least in the Euclidean region $0<x<1$.
Noting that the ladder diagrams appear with a factor of $(-1)^L$ per loop order,
this implies that $(-1)^L \Omega_{0}^{(L)}$ is positive for any $0<x<1$.
However, the fact that {\it all} signs within each of the above expressions are identical seems to
be a less trivial statement.

One more check that can be performed on the $\Omega_0^{(i)}$ is the limit $\phi \to 0$, corresponding
to $x \to 1$. In this limit the contribution of the ladders to the cusp anomalous dimension was
derived to all loop orders and to second order in $\phi$ in~\cite{Correa:2012nk} and reads
\begin{align}\label{eq:Gammalad}
\Gamma^{lad} &= \frac{1-\sqrt{\kappa+1}}{2} - \frac{\phi^2}{16} \, \kappa \left(\frac{1+\sqrt{\kappa+1}}{1+\kappa+2\sqrt{\kappa+1}}\right)
+ {\cal O}(\phi^4) \nonumber \\
& = \kappa \, \left[-\frac{1}{4} - \frac{\phi^2}{24}\right] + \kappa^2 \, \left[\frac{1}{16} + \frac{5\phi^2}{288}\right]
+ \kappa^3 \, \left[-\frac{1}{32} - \frac{43\phi^2}{3456}\right] + \kappa^4 \, \left[\frac{5}{256} + \frac{211\phi^2}{20736}\right] \nonumber \\
& \hspace*{10pt} +\kappa^5 \, \left[-\frac{7}{512} - \frac{4387\phi^2}{497664}\right] +\kappa^6 \, \left[\frac{21}{2048} + \frac{23545\phi^2}{2985984}\right] + {\cal O}(\kappa^7,\phi^4) \; ,
\end{align}
with $\kappa = \hat\lambda/\pi^2$. In order to verify this expansion we note that the ladder contribution to the cusp
anomalous dimension is given by
\begin{align}\label{eq:Gammaladomega}
\Gamma^{lad} &= - \sum\limits_{L\ge 1} \left(\frac{\lambda}{8\pi^2}\right)^L \, \left(-\frac{\xi}{2}\right)^L \, \Omega_0^{(L)} \; ,
\end{align}
and that in the limit we are interested in
\begin{align}
 \frac{\lambda}{8\pi^2} \, \xi & \rightarrow \frac{x}{2 \left(x^2-1\right)} \, \kappa \; .
\end{align}
Taking into account that $x = e^{i\phi}$ we expand~(\ref{eq:Gammaladomega}) to second order in $\phi$ and find
perfect agreement with~(\ref{eq:Gammalad}) through to six loops. In the next section, we will discuss the limit $x \to 0$.

\subsection{Simplifications in the $x \to 0$ limit}
The limit $x \to 0$ is interesting because it connects the velocity-dependent cusp anomalous dimension
discussed here with the light-like cusp anomalous dimension.\footnote{Since we have taken the
scaling limit we only have a subset of the usual diagrams. However, it is still interesting to discuss
their behaviour.} 

At four loops, taking the $x \to 0$ limit of eq. (\ref{omega4atLO}) leads to
\begin{align}\label{Omega4_xto0}
\Omega_{0}^{(4)}(x) \,\stackrel{x \to 0}{=}\,&  \frac{736}{315}\, \log^7 x + \frac{184 \pi^2}{45}  \, \log^5 x +16 \zeta_3 \, \log^4 x + \frac{76 \pi^4}{45}  \log^3 x + \left( \frac{32}{3} \pi^2 \zeta_3 + 40 \zeta_5 \right)  \log^2 x  \nonumber \\
&+  \left( \frac{34 \pi^6}{315} + 16 \zeta_3^2 \right) \,  \log x  + \left( \frac{8}{15} \pi^4 \zeta_3 + 4 \pi^2 \zeta_5 + 36 \zeta_7 \right) +\cO(x)\,. 
\end{align}
At five loops, we find
\begin{align}\label{Omega5_xto0}
\Omega_{0}^{(5)}(x) \,\stackrel{x \to 0}{=}\, &\;  - \frac{2144}{567} \, \log^9 x
- \frac{17152}{315} \, \zeta_{2} \, \log^7 x
- \frac{1472}{45} \, \zeta_{3} \, \log^6 x
- \frac{2896}{5} \, \zeta_{4} \, \log^5x \nonumber\\ &
- \left(\frac{736}{3} \, \zeta_{2} \, \zeta_{3} + \frac{304}{3} \, \zeta_{5}\right) \, \log^4 x
- (64 \, \zeta_{3}^2 + 1264 \, \zeta_{6}) \, \log^3 x \nonumber\\ &
- (912 \, \zeta_{3} \, \zeta_{4} + 288 \, \zeta_{2} \, \zeta_{5} + 168 \, \zeta_{7}) \, \log^2 x\nonumber\\ &
- \left(128 \, \zeta_{2} \, \zeta_{3}^2 + 160 \, \zeta_{3} \, \zeta_{5} + \frac{1240}{3} \, \zeta_{8}\right) \, \log x\nonumber\\ &
- \frac{32}{3} \, \zeta_{3}^3 - 144 \, \zeta_{4} \, \zeta_{5} - 204 \, \zeta_{3} \, \zeta_{6} - 
 72 \, \zeta_{2} \, \zeta_{7} - \frac{340}{3} \, \zeta_{9}+\cO(x)\,. 
\end{align}
Finally, at six loops, one obtains
\begin{align}\label{Omega6_xto0}
\Omega_{0}^{(6)}(x) \,\stackrel{x \to 0}{=}\,& \;   \frac{339008}{51975} \, \log^{11}x
+ \frac{339008}{2835} \, \zeta_{2} \, \log^9 x
+ \frac{4288}{63} \, \zeta_{3} \, \log^8 x\nonumber\\ &
+ \frac{12800}{7} \, \zeta_{4} \, \log^7 x
+ \left(\frac{34304}{45} \, \zeta_{2} \, \zeta_{3} + \frac{10688}{45} \, \zeta_{5}\right) \, \log^6 x\nonumber\\ &
+ \left(\frac{2944}{15} \, \zeta_{3}^2 + \frac{110944}{15} \, \zeta_{6}\right) \, \log^5 x
+ (5792 \, \zeta_{3} \, \zeta_{4} + 1376 \, \zeta_{2} \, \zeta_{5} + 528 \, \zeta_{7}) \,  \log^4 x\nonumber\\ &
+ \left(\frac{2944}{3} \, \zeta_{2} \, \zeta_{3}^2 + \frac{2432}{3} \, \zeta_{3} \, \zeta_{5} + \frac{80048}{9} \, \zeta_{8}\right) \, \log^3 x\nonumber\\ &
+ (128 \, \zeta_{3}^3 + 3792 \, \zeta_{4} \, \zeta_{5} + 7584 \, \zeta_{3} \, \zeta_{6} + 1152 \, \zeta_{2} \, \zeta_{7} + 664 \, \zeta_{9}) \, \log^2 x\nonumber\\ &
+ (1824 \, \zeta_{3}^2 \, \zeta_{4} + 1152 \, \zeta_{2} \, \zeta_{3} \, \zeta_{5} + 336 \, \zeta_{5}^2 + 672 \, \zeta_{3} \, \zeta_{7} + \frac{8292}{5} \, \zeta_{10}) \, \log x\nonumber\\ &
+ \frac{256}{3} \, \zeta_{2} \, \zeta_{3}^3 + 160 \, \zeta_{3}^2 \, \zeta_{5} + 612 \, \zeta_{5} \, \zeta_{6}
+ 432 \, \zeta_{4} \, \zeta_{7} + \frac{2480}{3} \, \zeta_{3} \, \zeta_{8} \nonumber \\ &
 + \frac{680}{3} \, \zeta_{2} \, \zeta_{9} + 372 \, \zeta_{11}+\cO(x)\,. 
\end{align}

It is worth noting that in~(\ref{Omega4_xto0})~--~(\ref{Omega6_xto0}) certain transcendental constants which correspond to Multiple Zeta Values~\cite{Blumlein:2009cf} having negative indices -- such as $\log(2)$ or ${\rm Li}_4(\frac{1}{2})$ --
do not appear. This becomes obvious from eqs.~(\ref{omega4atLO}),~(\ref{omega5atLO}), and~(\ref{omega6atLO}) at four, five, and six loops, respectively.
Moreover, eqs.~(\ref{Omega4_xto0})~--~(\ref{Omega6_xto0}) contain only single zeta values and products thereof. No Multiple Zeta Values
of depth 2 or higher appear up to six loops, although constants like $\zeta_{5,3}$ would be allowed in principle.

We would like to mention that there is a shortcut for obtaining the asymptotic limit,
without having to use the algorithm presented above.
It suffices to notice that to logarithmic accuracy as $x \to 0$, we can make the following replacement of the
integration kernel appearing e.g.\ in eq. (\ref{defHV0}),
\begin{align}
\frac{1}{w' + 1/x} - \frac{1}{w'+x} \longrightarrow - \frac{1}{w' +x} \; .
\end{align}
Next, rescaling all integration variables by $x$, we
see that one can write the result in the small $x$ limit at any loop order in terms
of HPLs with indices $0,-1$, and argument $1/x$.
The latter can be rewritten in terms of HPLs of argument $x$, and their small $x$ asymptotic behaviour can be
made manifest using algorithms implemented in \cite{Maitre:2005uu}.



 \section{NLO terms in large $\xi$ limit}
\label{sec:subleading}

\subsection{Triangle-ladder diagrams (b)}

We now wish to study the sum of the triangle-ladder diagrams shown in Fig.~\ref{fig:integrals_large_xi}(b)
in a similar way to LO. 
Let  $F$ now denote the sum of the diagrams of Figs.~\ref{fig:integrals_large_xi}(a,b), starting with $1$ (as at LO).
Then $F$ satisfies the Bethe-Salpeter equation of Fig.~\ref{fig:BetheSalpeter}, with the last term omitted.
(The last term will be discussed in the following section.)

Proceeding as at LO, we obtain the differential equation
\begin{align}\label{diffeqG}
\partial_{\sigma} \partial_{\tau} F(\sigma,\tau) =Q(\sigma, \tau) +  F(\sigma,\tau) P(\sigma,\tau)  \,.
\end{align}
Here the essential new feature is the appearance of $Q(\sigma,\tau)$. 
It arises from the first term in the second line of the r.h.s. of the equation shown in Fig.~\ref{fig:BetheSalpeter}.
It is given by the one-loop integral
\begin{align}
Q(\sigma,\tau) =& c\, \lambda \hat{\lambda}\, e^{(\sigma+\tau)}\, \int \frac{d^4 x_{1}}{i \pi^2} \frac{1}{x_{1}^2 (x_{1}-z_1 )^2 (x_{1}-z_{2})^2} 
=  c\, \lambda \hat{\lambda}\, \frac{ e^{(\sigma+\tau)}}{z_{12}^2} \Phi^{(1)}\left( \frac{z_1^2}{z_{12}^2}, \frac{z_{2}^2}{z_{12}^2} \right) \,,
\end{align}
where $z_1^{\mu} = e^\sigma p^{\mu}$ and $z_2^{\mu} = -e^\tau q^{\mu}$ are points along the Wilson line,
and $c=2/(8 \pi^2)^2$.
The function $\Phi^{(1)}$ is known analytically, and we will give a useful form for it later in this section.
Plugging in the expressions for $z^\mu_1,z^\mu_2$, we have
\begin{align}
Q(\tau,\sigma) =& c\, \lambda \hat{\lambda} \,  \frac{1}{\cosh(\tau-\sigma) + \cos \phi}
\; \Phi^{(1)}\left( 
\frac{ e^{\tau-\sigma}/2}{\cosh(\tau-\sigma) + \cos \phi}, 
\frac{ e^{\sigma-\tau}/2}{\cosh(\tau-\sigma) + \cos \phi} \right) \,.
\end{align}
Making the same ansatz as at LO, $F= \sum_n e^{-\Omega_{n}(\phi) y_{2}} \Psi_{n}(y_1,\phi)$, 
we obtain
\begin{align}\label{schroedinger2}
& \left[ -\partial_{y_1}^2 - \frac{\hat{\lambda}}{8 \pi^2} \frac{1}{(\cosh y_1 + \cos \phi)} + \frac{\Omega^2(\phi)}{4} \right] \Psi(y_1 ,\phi) 
= \nonumber \\ 
 & \qquad\qquad= 
 \, c\, \frac{\lambda\, \hat\lambda}{(\cosh y_1 + \cos \phi )}
\; \Phi^{(1)}\left( 
\frac{ e^{y_1}/2}{\cosh y_1 + \cos \phi}, 
\frac{ e^{-y_1}/2}{\cosh y_1 + \cos \phi} \right)\,.
\end{align}
We see that the essential new feature w.r.t.\ the LO case is the appearance of an inhomogeneous term.
It is important to realize that we would like to solve this equation to all orders in $\hat{\lambda}$, but only
to linear order in $\alpha = \lambda / \hat{\lambda}$, corresponding to the NLO case.

For simplicity of notation, let us abbreviate the potential by $- \hat \lambda V$ and the inhomogeneous term by
$\alpha \hat{\lambda}^2 \tilde{Q}$. Then we have
\begin{align}
\left[ -\partial_{y_1}^2 - \hat\lambda V(y_1,\phi) +\frac{\Omega^2(\phi)}{4} \right] \Psi(y_1, \phi) = \alpha \, \hat\lambda^2 \tilde{Q} \,.
\end{align}
Proceeding as in the homogeneous case and setting $\Psi = e^{-\Omega/2 y_1} \eta$
we have
\begin{align}
-\partial^2_{y_1}\eta + \Omega \partial_{y_1} \eta - \hat\lambda V \eta = \alpha\, e^{+\Omega /2 \, y_1} \, \hat{\lambda}^2\, \tilde{Q} \,.
\end{align}
Recall that at $\alpha=0$, this is just the equation for the ladder diagrams, which we already solved.
We need the solution to order $\alpha$. We can expand
\begin{align}
\eta = \eta_{\rm ladders} + \alpha \, \eta_{\alpha}\,,\quad \Omega = \Omega_{\rm ladders} + \alpha \, \Omega_{\alpha} \,,
\end{align}
to obtain, at order $\alpha$,
\begin{align}\label{schroedingeralpha}
- \partial_{y_1}^2 \eta_{\alpha} + \Omega_{\rm ladders} \partial_{y_1} \eta_{\alpha} - \hat{\lambda} V \eta_{\alpha} = e^{y_1 \Omega_{\rm ladders}/2} \hat{\lambda}^2 \tilde{Q} - \Omega_{\alpha} \eta_{\rm ladders}' \,.
\end{align}
As before, $\Omega$ is obtained by requiring that $\Psi'(y_{1})$ vanishes at $y_{1}=0$.
Therefore we have
\begin{align}
\Omega = 2 \partial_{y_1} \log \eta {|}_{y_1 =0}\,.
\end{align}
At order $\alpha$, this gives
\begin{align}\label{omega_alpha}
\Omega_{\rm \alpha} =& \, 2 \partial_{y_1} \left( \frac{\eta_{\alpha}}{\eta_{\rm ladders}} \right) {|}_{y_1 =0} \,.
\end{align}
In summary, we have arrived at a differential equation, eq. (\ref{schroedingeralpha}), together with (\ref{omega_alpha}),
for the contribution of  the triangle-ladder diagrams shown in Fig.~\ref{fig:integrals_large_xi}(b).

We will now explain how to solve these equations to any order in $\hat\lambda$.
First of all, it is clear that we can integrate order by oder in $\hat\lambda$ just as we did at LO.
The main question is whether we can express the resulting wavefunction at each order in terms
of the same set of iterated integrals as in the previous section.
We will now show that this is indeed the case, and in fact is true also for a more general
class of diagrams.

The new feature of eq. (\ref{schroedingeralpha}) is the appearance of $\tilde{Q}$, so we need
to analyze whether integrals over $\tilde{Q}$ will be of the same form as at LO.
An example will suffice to see that this is indeed the case. Consider expanding to order $\hat\lambda^2$.
Then $\eta^{(1) \, \prime}(w,x)$ is given by an integral of the form 
\begin{align}\label{eqintrgralb1}
\int_{-\log w}^{\infty} \,  \frac{d y_1}{(\cosh y_1 + \cos \phi )}
\; \Phi^{(1)}\left( 
\frac{ e^{y_1}/2}{\cosh y_1 + \cos \phi}, 
\frac{ e^{-y_1}/2}{\cosh y_1 + \cos \phi} \right) \; .
\end{align}
We will now make use of the fact that $\Phi^{(1)}$ is a function with very special properties.
In fact, this allows us to immediately make a generalization where $\Phi^{(1)}$ is replaced by $\Phi^{(n)}$.
This function is given by a beautiful formula \cite{Isaev:2003tk},
\begin{align}
\Phi^{(n)}(x,y) = \frac{1}{\sqrt{(1-x-y)^2 - 4 x y}} \, \tilde{\Phi}^{(n)}(x,y) \,,
\end{align}
where
\begin{align}\label{isaev}
\tilde{\Phi}^{(L)}(x,y) =  \sum_{f=0}^{L} \frac{(-1)^f (2L-f)!}{L! f! (L-f)!} \log^{f}(z_1 z_2) \left[ {\rm Li}_{2L-f}(z_{1}) - {\rm Li}_{2 L -f}(z_2 ) \right]\,,
\end{align}
and
\begin{align} \label{defz1z2}
x = z_1 z_2 \,,\qquad y =  (1-z_1 )(1-z_2 ) \,.
\end{align}
Changing variables to $w'=e^{-y_1}$ and $x=e^{i \phi}$,
eq. (\ref{eqintrgralb1}) becomes, up to a trivial normalization factor,
\begin{align}
\int_0^w \, \frac{dw'}{w'} \, \tilde{\Phi}^{(1)}\left( \frac{1}{w'^2 +2 w' \cos \phi +1}, \frac{w'^2}{w'^2 +2 w' \cos \phi +1} \right) \,.
\end{align}
Inspection shows that the variables defined in (\ref{defz1z2}) are given by
\begin{align}
z_{1} = \frac{x}{x+w'} \,,\qquad z_{2} = \frac{1}{1+x w'}\,.
\end{align}
Furthermore, the functions above can be defined using only
iterative integrals corresponding to symbols $z_1, z_2, 1-z_1, 1-z_2$.
It is easy to verify that the latter factorize over $x,w,w+x,1+w x$,
and hence are contained in the function class discussed in the previous section.
This  implies that we can again perform all iterated integrals within the
set of polylogarithms defined by the same integration kernels/symbols as
in the homogeneous case, and therefore allowing for an algorithmic 
solution of this problem.

We note that there is an obvious generalization to a class of diagrams 
where $\Phi^{(1)}$ is replaced by $\Phi^{(n)}$, see Appendix A of ref. \cite{Correa:2012nk}.
The perturbative solution for that class of diagrams can be done in the same
way as explained above.

\subsection{H-exchange diagrams (c)}
The diagrams with H-exchange of  Fig.~\ref{fig:integrals_large_xi}(c) were analyzed in ref. 
\cite{Bykov:2012sc} for the quark-antiquark potential. It was found that the Bethe-Salpeter
equation in that case contains a new term of the form
\begin{align}\label{newtermc}
\int_0^{\infty} du \int_0^\infty dv\, e^{-\frac{\Omega_{0}}{2} (u+v)} \, f(u,v;y_1) \Psi(y_1 - u+v) \,,
\end{align}
so that one has a linear integro-differential equation for $\Psi$. 
Their analysis can be adapted to the present case of general $\phi$, 
with $f$ now depending on $\phi$.

Although such an equation may seem complicated, it simplifies considerably when solving
it in the small $\alpha = \lambda/\hat{\lambda}$ limit. The reason is that the kernel, the $H$-exchange 
diagram is already of order $\alpha$, so that we only need the wavefunction at order $\alpha^0$.
In other words, the problem reduces to a differential equation for the wavefunction at order $\alpha$,
with an inhomogeneous term. This is exactly the case we studied in the previous section.

Having said this, the main difficulty lies in the computation of the H insertion, and
in integrating it when iteratively solving for the wavefunction.
From the discussion above it is clear that we need to understand
how to carry out the H-shaped and similar integrals.
Let us therefore start with the basic three-loop integral, which has one H-exchange, 
and no additional rungs. It is given by
\begin{align}
 \int_0^\infty ds_2 \int_0^{s_2} ds_1 \int_0^\infty dt_2 \int_0^{t_2} dt_1 \, f(-s_1 p^{\mu} , -s_2 p^{\mu}; t_1 q^{\mu} , t_2 q^{\mu} ) \,.
\end{align}
Note that strictly speaking we should introduce IR and UV regulators for this integral, 
but since we are only interested in extracting the overall divergence, the details of the cutoffs are
not very important.
For the same reason, the H-shaped subintegral can be defined in exactly four dimensions,
\begin{align}
f(x_1,x_2,x_3,x_4)=&  (\partial_{1}+ \partial_{4})^2 \,  h(x_1,x_2;x_3,x_4) \,, \label{eq:deff}\\
h(x_1,x_2;x_3,x_4) =& \int\frac{d^{4}x_{5} d^{4}x_{6}}{(i \pi^2)^2} \, \frac{1}{x_{15}^2 x_{25}^2 x_{36}^2 x_{46}^2  x_{56}^2} \label{defh}\,.
\end{align}
Eq.~(\ref{eq:deff}) defines the function $f$. Although this is a two-loop integral,
$f$ reduces to one-loop integrals thanks to differential equations it satisfies. 
We review these differential equations in Appendix~A. Remarkably, they allow us to express $f$ in terms of the one-loop
function $\Phi^{(1)}$, the same function that appeared already in integral class (b). Explicitly, we have
\begin{align}\label{result_f}
 \tilde{f} \;=& \;x^2_{24} (x^2_{12} + x^2_{23} - x^2_{31}) \,
\Phi^{(1)}\left( \frac{x^2_{12}}{x^2_{13}}, \frac{x^2_{23}}{x^2_{13}}\right) + 
   x^2_{13} (x^2_{12} + x^2_{14} - x^2_{24}) \, \Phi^{(1)}\left( \frac{x^2_{12}}{x^2_{24}}, \frac{x^2_{14}}{x^2_{24}}\right) \nonumber\\
   & + 
   x^2_{24} ( x^2_{14} + x^2_{34}-x^2_{13} ) \,  \Phi^{(1)}\left( \frac{x^2_{34}}{x^2_{13}}, \frac{x^2_{14}}{x^2_{13}} \right) 
   + x_{13}^2   ( x^2_{23}  + x^2_{34} -x^2_{24}) \, \Phi^{(1)}\left( \frac{x^2_{34}}{x^2_{24}}, \frac{x^2_{23}}{x^2_{24}} \right) \nonumber\\
   & 
   + (  x^2_{13} x^2_{24} - x^2_{14} x^2_{23}-   x^2_{12} x^2_{34}) \,
   \Phi^{(1)}\left( \frac{x^2_{12} x^2_{34}}{x^2_{13} x^2_{24}}, \frac{x^2_{14} x^2_{23}}{x^2_{13} x^2_{24}} \right) \,,
\end{align}
where $\tilde{f} = (x^2_{12} x^2_{13} x^2_{24} x^2_{34}) f$.
This formula will be very convenient when discussing the strong coupling limit.

After this digression on $h$, we can proceed to extract the overall divergence
and compute the H-exchange integral. Changing variables according to $s_{1} = x_1 s_2 , t_1 = x_2 t_2$, and
$s_2 = z \rho, t_2 = \rho \bar{z}$, where $\bar{z}=1-z$, and using that $h$ scales as $1/x^4$, we find
\begin{align}
 \int_0^{\infty} \frac{d\rho}{\rho} \, H^{(3)}\,,
\end{align}
where
\begin{align}\label{eqforH3}
H^{(3)} =  \int_0^1 \, dz \, dx_1 \, dx_2 \, f(-x_1 z p^{\mu},  -z p^{\mu}; x_2 \bar{z} q^{\mu}, \bar{z} q^{\mu}) \,.
\end{align}
Note that by assumption $H^{(3)}$ is finite (i.e.\ the original integral only had an overall UV divergence).
However, when carrying out the integration in (\ref{eqforH3}), care is required, because the finiteness
is not necessarily true for individual terms appearing in (\ref{result_f}). This small problem can
be avoided by introducing an auxiliary regulator.
With the above parametrization, we have
\begin{align}
x_1^\mu  = -x_1 z p^\mu \,,\quad x_2^\mu = -z p^\mu\,,\quad x_3^\mu = x_2 \bar{z} q^\mu\,,\quad x_4^\mu = \bar{z} q^\mu \,,
\end{align}
and using $p^2=q^2=1\,,  p\cdot q = \cos \phi$, we have
\begin{align}
x_{12}^2 =& \bar{x}_1^2 z^2 \,,\quad x_{23}^2 = z^2 + x_2^2 \bar{z}^2 + x_2 z \bar{z} 2 \cos \phi \,,
\end{align}
and so on.

In summary, we found a finite parameter integral, where the number of integrations equals the
expected degree of the function. Just as for integral class (b), higher orders can be obtained by iteration.
However, it is not yet clear that the same class of functions will be sufficient to evaluate these integrals.
Explicit results at three loops motivate that it might be.
We leave this question for future work, and close this section by remarking that
formula (\ref{isaev}) will certainly be very useful when trying to evaluate this integral
and similar integrals appearing in the iterative solution.



\section{Strong coupling limit at LO and NLO}
Here we discuss the strong coupling limit of the
Bether-Salpeter equations. 
In this limit, the calculation of the ground state energy
becomes almost trivial. 
It is straightforward to extend the analysis of ref. \cite{Bykov:2012sc},
which was done in the anti-parallel lines limit $\phi \to \pi$, to any angle.

\subsection{Strong coupling limit of Bethe-Salpeter equation}

Let us start by discussing diagrams of type (c).
First of all, we notice that as in \cite{Bykov:2012sc}, the Bethe-Salpeter equation
for this class of diagrams simplifies dramatically in the strong coupling limit.
The reason is that for $\Omega_0 \sim \sqrt{\lambda} \gg 1$, the region
of small $u,v$ will give the dominant contribution to the integral in eq. (\ref{newtermc}).
This implies that the wavefunction $\Psi(y_1)$ can be pulled out of the integral,
with the coefficient being an effective potential.
This argument also works for the angle-dependent cusp Wilson loop.

We therefore need to compute the effective potential for general angles.
Although the function $h$ is not known analytically, its derivative $f$ is known,
as we saw in the previous section.

We need the function $f(x_1,x_2;x_3,x_4)$ in the limit where $x_{1} \rightarrow x_{4}$ 
and $x_{2} \rightarrow x_{3}$. Let us parametrize this limit by $x_{14}^2 =  u^2, x_{23}^2 =  v^2, x_{12}^2 = x_{24}^2 =x_{13}^2 =x_{34}^2 = 2 \cosh y_1 + 2 \cos \phi$, with $u,v$ small.
Plugging these values into eq. (\ref{result_f}), it turns out we only need the following limit of $\Phi^{(1)}$,
\begin{align}
\Phi^{(1)}(1,\eps) = - \log \eps+ 2 + \cO(\eps) \; .
\end{align}
Using this limit, we obtain
\begin{align}
f \longrightarrow - 4  \frac{ u^2 \log {u}  + v^2 \log {v}}{ (2 \cosh y_1 + 2 \cos \phi)^{3}} + \cO(u^2,v^2)\,.
\end{align}
We see that this is a generalization of eq. (5.2) of \cite{Bykov:2012sc}
to general angles. One could also use eq. (\ref{result_f}) to compute 
higher order terms in the expansion.

This means that the correct effective potential for the general angle case is
obtained by replacing each $(x^2+1)$ terms in (5.3) of \cite{Bykov:2012sc}
by $(2 \cosh y_1 + 2  \cos \phi )$ for the cusped Wilson loop.
Then we have a Schr\"{o}dinger equation 
\begin{align}
\left[ -\partial_{y_1}^2 + V_{\rm eff}(y_1) + \frac{ \Omega^2}{4} \right] \Psi(y_1) = 0
\end{align}
where the correction to the effective potential comes from the integral
\begin{align}
V_{\rm eff} |_{\lambda \hat{\lambda}^2} \sim \int_0^\infty \int_0^{\infty} du\, dv \, e^{-\frac{\Omega}{2} (u+v)} \, f(u,v) \,.
\end{align}
Explicitly, we have 
\begin{align}
V_{\rm eff} = -\frac{\hat{\lambda}}{4 \pi^2 (2 \cosh y_1 + 2 \cos \phi )} + \frac{ \lambda \hat{\lambda}^2  \log \Omega }{2 \pi^6 \Omega^4  (2 \cosh y_1 + 2 \cos \phi )^3} \,.
\end{align}
At strong coupling, we can focus on $\hat\lambda \gg1, y_1 \ll1$, with $\hat\lambda (y_1)^{1/4}$ fixed. In that regime 
the leading term of the Schr\"{o}dinger equation is
\begin{align}
 V_{\rm eff}(y_1 = 0) +  \frac{ \Omega_0^2}{4} = 0 \,.
\end{align}
From this we obtain for the ground state energy,
\begin{align}\label{gammacusp_fieldtheory}
\Gamma^{(a)+(c)} = - \Omega_0 = - \frac{\sqrt{\hat{\lambda}}}{2 \pi \cos \frac{\phi}{2}} \left[ 1- \frac{1}{2} \frac{ \lambda}{ \hat{\lambda}}   \log \frac{\hat\lambda}{\lambda} + \cO\left(  \frac{\lambda}{\hat{\lambda}}   \right) \right] 
\end{align}
Here the superscript indicates that this is the contribution from the integrals
shown in Figs.~\ref{fig:integrals_large_xi}(a),(c).

Let us now discuss the integrals of Fig.~\ref{fig:integrals_large_xi}(b).
Here we obtained a Schr\"{o}dinger equation with an inhomogeneous term (note that there $\alpha = {\lambda/\hat{\lambda}}$)
that is not multiplied by the wave function.
The latter fact  suggests to us that the contribution of this class of diagrams at strong coupling
will not be given by an exponential factor of the type seen for integral classes (a) and (c).
If one assumes the absence of 
contributions of integral class (b) at strong coupling, as we will do in the following, 
then (\ref{gammacusp_fieldtheory}) is
the full answer at LO and NLO in the scaling limit.

Let us now compare this against the corresponding quantity computed in string theory.

\subsection{Scaling limit of the string theory result}
The leading term (and first subleading term as well) in the $1/\sqrt{\lambda}$ expansion 
at strong coupling
has been computed using string theory in ref. \cite{Drukker:2011za}. 
It is straightforward to expand their result in the large $\hat{\lambda}$ 
limit that we are interested in. For the LO, this was already
done in ref. \cite{Correa:2012nk}. 

It is easy to take the scaling limit of the formula for $\Gamma$ given in ref. \cite{Drukker:2011za}.
The details of this calculation are presented in Appendix~B.
We find
\begin{align}\label{gamma_strong}
\Gamma = - \frac{\sqrt{\hat\lambda}}{2 \pi \cos \frac{\phi}{2}}  
\left[  1 -  \frac{ 1}{2} \frac{\lambda}{\hat{\lambda}} \log \frac{\hat{\lambda}}{\lambda} + \cO\left(  \frac{\lambda}{\hat{\lambda}} \right) 
\right] \,.
\end{align}
As a consistency check, we can take the limit $\phi = \pi-\delta$, $\delta \to 0$,
where we expect to find the quark-antiquark potential $V$. More precisely, $\Gamma \sim - 1/\delta\, V$, 
and indeed we find agreement with eq. (5.4) of \cite{Bykov:2012sc}.

Let us compare eq. (\ref{gamma_strong}) to the diagram calculation performed in the previous
subsection. Comparing to eq. (\ref{gammacusp_fieldtheory}), we find perfect agreement.
Recall that in principle there could also be a contribution from integrals of type (b) not accounted for
in eq. (\ref{gammacusp_fieldtheory}), but we argued that this is not the case based on the structure of
the Bethe-Salpeter equation for these integrals.
Under this assumption, we see that there is a perfect match between the field theory calculation in
the scaling limit, and the string theory calculation. As pointed out in \cite{Correa:2012nk}, this agreement was not
guaranteed due to potential order of limits issues.



\section{Discussion and conclusion}\label{sec:conc}

In this paper we further studied the scaling limit of the cusp anomalous dimension
introduced in \cite{Correa:2012nk}, in several ways. 

In the first part of the paper, working at LO we showed that the perturbative solution
at weak coupling can be expressed at any loop order in terms of
harmonic polylogarithms, and outlined a corresponding algorithm.
As illustration, we reproduced the three-loop result of  \cite{Correa:2012nk}
and computed the four-, five-, and six-loop results, which are new.
We also provide a shortcut for obtaining the $x \to 0$ asymptotics,
which corresponds to the
light-like limit of the edges of the Wilson loop.

We observed interesting features of these results.
We find that, at least up to six loops, they can be written in
terms of a reduced class of harmonic polylogarithms, with
indices $0$ and $1$ only, when choosing $x^2$ as argument
(this feature was already noted in  \cite{Correa:2012nk} up to three loops.).
Moreover, in the $x \to 0$ limit, again up to six loops, we find
that the resulting asymptotic expansion can be expressed 
in terms of linear combinations of products of single zeta values
only. Other constants such as $\log(2)$, or multiple
zeta values of higher depth were not needed.
This is especially interesting in the context of the BES
equation for the closely related light-like cusp anomalous dimension,
which has the same property \cite{BES}. 

It would be very interesting if one could prove 
these properties. Such a proof would likely shed more light
on the structure of the cusp anomalous dimension.

In the second part of the paper, we extended the analysis of \cite{Correa:2012nk}
to NLO order. 
The new feature of the equations is the appearance of an 
inhomogeneous term. (A similar analysis was recently done
for the quark-antiquark potential \cite{Bykov:2012sc}). This term does not alter the
perturbative solution, however, and we were able to apply the
same strategy as at LO.
We showed how to compute these contributions
systematically in perturbation theory.
For one class of integrals, we provided an algorithmic
solution at any loop order in terms of harmonic polylogarithms.
For the second class of integrals, we showed how to obtain
the solution in terms of iterated integrals of simple functions. 
We left the question of whether the latter can be expressed
in terms of HPLs for future work.

Finally, we discussed the strong coupling limit of the 
equations.
We computed the logarithmically enhanced terms at NLO, and 
found agreement between the field theory and the string theory
calculation. This generalizes the calculation of \cite{Bykov:2012sc} to
any angle.
Using our formulas, it should be possible to compute the non-logarithmic terms
at NLO as well. We leave this for future work as well.

In ref.~\cite{Correa:2012nk} the zero angle case was studied, where the Schr\"{o}dinger
potential becomes the exactly solvable P\"{o}schl-Teller potential. It would be
interesting to extend this analysis to NLO, where the equation is modified by an
inhomogeneous term, as discussed in the present paper.


Our approach also suggests a general strategy for the computation
of the cusp anomalous dimension, or related quantities. At a given
loop order, there are two sets of contributions.
First, there are a number of integrals that have an overall UV divergence.
These diagrams are the ``seed'' of the Bethe-Salpeter equations and
have to be computed. They correspond to the 
most complicated part of the calculation. However, the fact that they
have no subdivergences allows one to extract the overall divergence
easily, so that one is left with the calculation of a finite quantity. The latter is sometimes
related to four-dimensional integrals. This observation allowed for
example for the computation of an infinite class of integrals contributing
to the cusp anomalous dimension in ref.~\cite{Correa:2012nk}.
Second, there are diagrams that do have subdivergences.
For these contributions, the resummation technique via the Bethe-Salpeter
equation is very useful, as it automatically takes into account the 
non-Abelian exponentiation. Although these contributions typically
give the most complicated contributions as far as the functions involved
are concerned~\cite{Correa:2012nk}, the latter have their origin in
simple iterations of diagrams of the first type.

Although our analysis did not rely on the planar limit,
non-planar contributions to $\Gamma_{\rm cusp}$ appear
only at four loops, or at higher subleading terms in the scaling limit.
It would be very interesting to compute the first non-planar corrections.
We expect that many observations about the calculation of loop integrals,
especially the comments for extracting overall divergences and 
using four-dimensional integrals, will be useful in related problems
as well, e.g. as the non-planar integrals discussed in ref. \cite{Dixon:2009ur}.

Our approach can also be extended beyond the NLO.
We remark that this does not require any Feynman
graph calculations, as the integrand for the planar 
Wilson loop can be obtained from a soft limit of the 
integrand of a four-particle scattering amplitude \cite{Henn:2010bk,Correa:2012nk}.
The latter can be obtained through on-shell recursion
relations in principle to any loop order. 
We give examples of this procedure in Appendix~C.
We hope that this all-loop knowledge of the Wilson loop 
integrand gives a good starting point for analyzing
further the properties that we have observed in 
this paper.

Finally, the scaling limit discussed here might be useful for simplifying the
TBA equations of refs. \cite{IntegrabilityPaper,Drukker:2012de}. 
It would also be interesting if those equation could shed light on some
of the observations about the perturbative properties of $\Gamma_{\rm cusp}$ 
that we have made here.


\section*{Acknowledgments}
It is a pleasure to thank N.~Arkani-Hamed, F.~Brown, S.~Caron-Huot, A.~Sever, and A.~Zhiboedov for helpful conversations.
JMH was supported in part by the Department of Energy grant DE-FG02-90ER40542.
TH is supported by the Helmholtz Alliance ``Physics at the Terascale''. 


\appendix

\section{Differential equations for two-loop integral $h$}

Here we give the differential equations for the two-loop integral $h$ of eq. (\ref{defh}).

The non-trivial differential equation has been written down in ref.~\cite{Eden:1999kh} and in eq. (A.7) of ref.~\cite{Beisert:2002bb}.
The differential operator we have can be related to the one of eq. (A.7) of~\cite{Beisert:2002bb} by using translational invariance ($\sum_{i=1}^{4} \partial_i  =0$), up to trivial pieces proportional to $\square_{i}$, where we can use the Laplace equation.
Note that all occurring terms can be written in terms of $\Phi^{(1)}$, thanks to eq.~(A.5) below.

More explicitly, we have
\begin{align}
-2 (\partial_1 + \partial_4)^2 = - (\partial_1 - \partial_2)\cdot (\partial_2 -\partial_3) - (\partial_1 + \partial_2 )^2 + \square_1 + \square_2 + \square_3 + \square_4 \,.
\end{align}
We have, using  eq. (A.7) of ref. \cite{Beisert:2002bb}, up to overall factors,
\begin{align}\label{diffeqh1}
(\partial_1 - \partial_2)\cdot (\partial_2 -\partial_3) h =& \frac{1}{x_{12}^2 x_{34}^2} 
\Big[ (x_{13}^2 x_{24}^2 - x_{14}^2 x_{23}^2) X_{1234}+ (x_{14}^2-x_{13}^2) X_{134} 
  \nonumber \\
&
-(x_{24}^2-x_{23}^2) X_{234}
+(x_{32}^2-x_{31}^2) X_{312} 
- (x_{42}^2-x_{41}^2) X_{412} \Big]\,.
\end{align}
and, from the Laplace equation,
\begin{align}\label{diffeqh2}
\left[- (\partial_1 + \partial_2 )^2 + \square_1 + \square_2 + \square_3 + \square_4 \right] h =&
-X_{1234}+\frac{1}{x_{12}^2} (X_{134} +X_{234}) + \frac{1}{x_{34}^2} (X_{123}+X_{124}) \,,
\end{align}
where
\begin{align}\label{X1234}
X_{1234} =& \int \frac{d^{4}x_{i}}{i \pi^2} 
\frac{1}{x_{1i}^2 x_{2i}^2 x_{3i}^2 x_{4i}^2} 
= \frac{1}{x_{13}^2 x_{24}^2} 
\Phi^{(1)}\left( \frac{x_{12}^2 x_{34}^2}{x_{13}^2 x_{24}^2} , 
\frac{x_{14}^2 x_{23}^2}{x_{13}^2 x_{24}^2} \right) \,, \\
X_{123} =& \int \frac{d^{4}x_{i}}{i \pi^2} \frac{1}{x_{1i}^2 x_{2i}^2 x_{3i}^2}  = \frac{1}{x_{13}^2 } \Phi^{(1)}\left( \frac{x_{12}^2 }{x_{13}^2} , \frac{x_{23}^2 }{x_{13}^2 } \right) \,.\label{X123}
\end{align}
Combining differential equations (\ref{diffeqh1}) and (\ref{diffeqh2}), and plugging in 
(\ref{X1234}) and (\ref{X123}), we find eq. (\ref{result_f}) given in the main text.

\section{Details of the strong coupling calculation}
Here we show the details of the expansion of the string theory answer for $\Gamma$ in the
scaling limit. The result of \cite{Drukker:2011za} for $\Gamma$ is parametrized by two parameters $p$ and $q= -i r\,, r>0$,
which are implicitly defined through the angles $\phi$ and $\theta$, in the following way
\begin{align}
\theta =& \frac{2 b q}{\sqrt{b^4 + p^2}} K(k^2) \,, \\
\phi =& \pi- 2 \frac{p^2}{b \sqrt{b^4+p^2}} \left[ \Pi\left(\frac{b^4}{b^4+p^2}, k^2 \right) - K(k^2) \right] \,, 
\end{align}
where
\begin{align}
b^2 =& \frac{1}{2} \left( p^2-q^2+\sqrt{(p^2-q^2)^2+4 p^2} \right) \,,\\
k^2 =& \frac{b^2 (b^2-p^2)}{b^4+p^2} \,.
\end{align}
In terms of these variables, we have
\begin{align}
\Gamma =& \frac{\sqrt{\lambda}}{2 \pi} \frac{2 \sqrt{b^4+p^2}}{b p} \left[ \frac{(b^2+1) p^2}{b^4+p^2} K(k^2) -E(k^2) \right] \,,
\end{align}
where $E,K$ and $\Pi$ are complete elliptic integrals, 
\begin{align}
E(k^2) =& \int_0^\frac{\pi}{2} dt \frac{1}{\sqrt{1-k^2 \sin^2 t}} \,,\\
K(k^2) =& \int_0^\frac{\pi}{2} dt \sqrt{1-k^2 \sin^2 t} \,,\\
\Pi(a^2,b^2) =& \int_0^\frac{\pi}{2} dt \frac{1}{(1-a^2 \sin^2 t)  \sqrt{1-b^2 \sin^2 t} }\,.
\end{align}
The scaling limit $i\theta \gg 1$ is reached by letting $p \to 0$.
We see that we require the leading and subleading divergences of $\Pi$ in the limit where
\begin{align}
\lim_{\eps \rightarrow 0} \Pi(1-a \eps ,1-\eps b) & =  \frac{1}{\eps} \, \frac{ \pi -2 \arcsin \frac{ \sqrt{a}}{\sqrt{b}}}{2 \sqrt{a} \sqrt{b-a}} - \frac{1}{4}  \log(\eps) \nonumber \\
&- \frac{ \sqrt{a}\left(\pi -2 \arcsin \frac{ \sqrt{a}}{\sqrt{b}}\right)}{4 \sqrt{b-a}} -\frac{1}{4} \log(b)
+\frac{1}{4} +\log(2) +\cO(\eps) \,.
\end{align}

In this way, we obtain, at leading order in $\sqrt{\lambda}\gg1$,
\begin{align}\label{resultgamma1}
\Gamma = - \frac{r \sqrt{\lambda}}{ \pi p} \left[ 1 + p^2 \log p  \frac{ (1+r^2)}{2 r^4}   \right]+ \cO(p)
\end{align}

We now convert  $r$ and $p$ to their expressions in terms
of $\theta$ and $\phi$. So we need the expansions of the latter to the necessary order in $p$.
We find
\begin{align}
e^{i \theta/2} =& \frac{1}{p}  \, 4 \frac{r^2}{\sqrt{1+r^2}} + p \log p \, \frac{3 + r^2}{r^2 \sqrt{1+r^2}} + \cO(p) \,,  \label{expansion1}\\
\phi =& 2 \arcsin \frac{1}{\sqrt{1+r^2}} - \frac{1}{r^3} p^2 \log p + \cO(p^2) \label{expansion2}\,,
\end{align}
We can see that this is in agreement with equation (\ref{gamma_strong}) quoted in the main text.

\section{Relation between integrals for four-particle scattering amplitude and cusp anomalous
dimension}

{}
\FIGURE[h!]{
\includegraphics[width=0.7\textwidth]{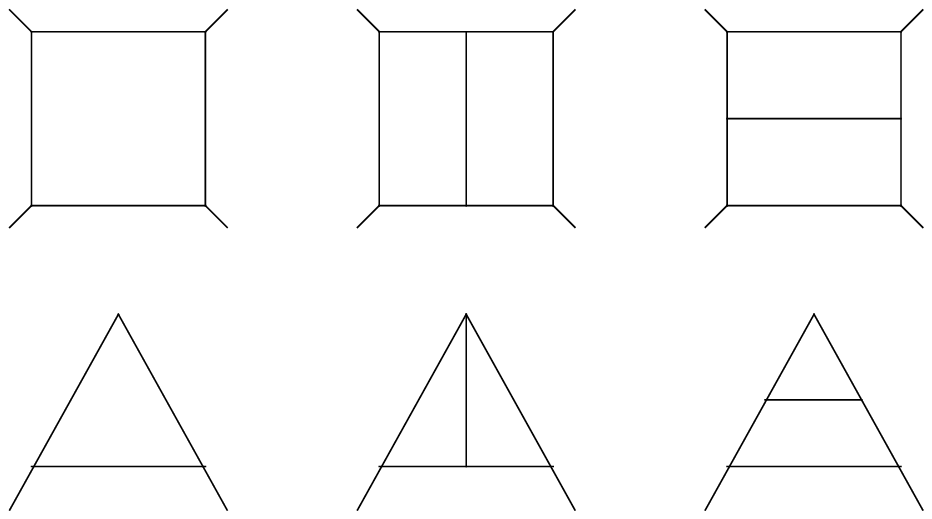}
\caption{
Relation between integrals of the four-point amplitude (first line) and Wilson line integrals (second line) at one and two loops.}
\label{fig:4ptWilsonloop1}
}

{}
\FIGURE[h!]{

\vspace*{1.0cm}

\includegraphics[width=0.65\textwidth]{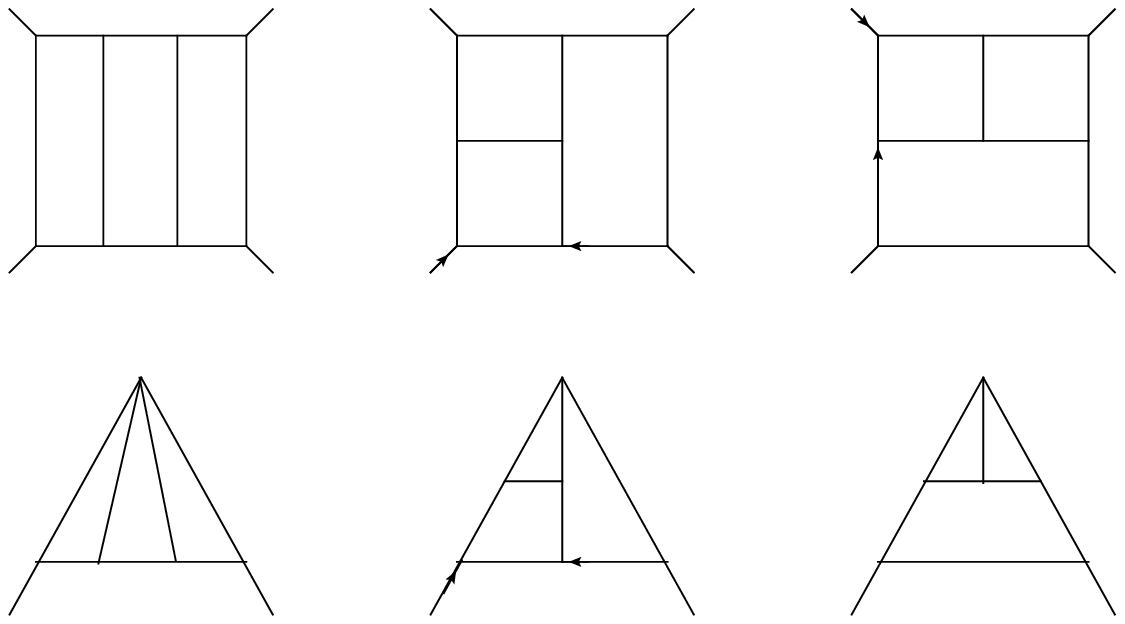}

\includegraphics[width=0.40\textwidth]{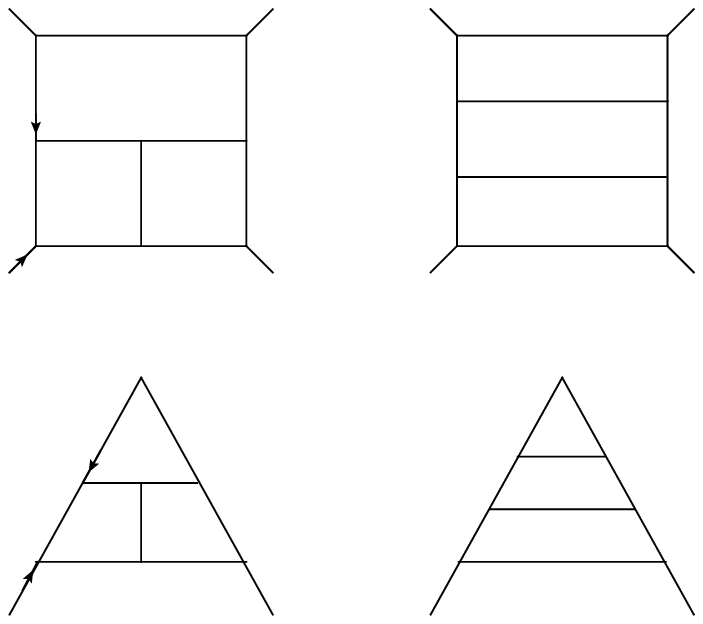}
\caption{
Relation between integrals of the four-point amplitude (first and third line) and Wilson line integrals (second and fourth line) at three loops.
Arrows denote internal numerator factors $(p^\mu + q^\mu)^2$, where $p^\mu$ and  $q^\mu$ are the momenta flowing along the lines with arrows.}
\label{fig:4ptWilsonloop2}
}

In Figs.~\ref{fig:4ptWilsonloop1} and~\ref{fig:4ptWilsonloop2} we illustrate the relation between the integrals contributing to the four-particle scattering amplitude (odd lines)
and the integrals contributing to the cusp anomalous dimension (even lines) to three loops.
The integrals occurring at one and two loops are shown in Fig.~\ref{fig:4ptWilsonloop1}.
The three-loop integrals are shown in Fig.~\ref{fig:4ptWilsonloop2}.
The reason for this relation \cite{Henn:2010bk,Correa:2012nk} is essentially exact dual conformal symmetry \cite{Alday:2009zm},
together with the fact that massive scattering amplitudes in soft limits are related to Wilson loops \cite{Korchemsky:1991zp}.

\section{The six-loop function $\Omega_0^{(6)}$}

Here we present the six-loop result $\Omega_0^{(6)}$ which we relegated from the main text to this Appendix.
Again, all HPLs are understood to have argument $x^2$.

{\allowdisplaybreaks
\begin{align}\label{omega6atLO}
\Omega_{0}^{(6)}(x) =&\; \frac{256}{3} \, \zeta_{2} \, \zeta_{3}^3 + 160 \, \zeta_{3}^2 \, \zeta_{5} + 612 \, \zeta_{5} \, \zeta_{6} + 
 432 \, \zeta_{4} \, \zeta_{7} + \frac{2480}{3} \, \zeta_{3} \, \zeta_{8} + \frac{680}{3} \, \zeta_{2} \, \zeta_{9} + 
 372 \, \zeta_{11} \nonumber \\ &
 + 912 \, \zeta_{3}^2 \, \zeta_{4} \, H_{0} + 
 576 \, \zeta_{2} \, \zeta_{3} \, \zeta_{5} \, H_{0} + 168 \, \zeta_{5}^2 \, H_{0}+ 
 336 \, \zeta_{3} \, \zeta_{7} \, H_{0} + \frac{4146}{5} \, \zeta_{10} \, H_{0} \nonumber \\ &
 + \frac{320}{3} \, \zeta_{3}^3 \, H_{2} + 1440 \, \zeta_{4} \, \zeta_{5} \, H_{2} + 
 2040 \, \zeta_{3} \, \zeta_{6} \, H_{2} + 720 \, \zeta_{2} \, \zeta_{7} \, H_{2} + 
 \frac{3400}{3} \, \zeta_{9} \, H_{2} \nonumber \\ &
  + 1024 \, \zeta_{2} \, \zeta_{3}^2 \, H_{3} + 
 1280 \, \zeta_{3} \, \zeta_{5} \, H_{3} + \frac{9920}{3} \, \zeta_{8} \, H_{3} + 
 9744 \, \zeta_{3} \, \zeta_{4} \, H_{4} + 3360 \, \zeta_{2} \, \zeta_{5} \, H_{4} \nonumber \\ &
 + 2664 \, \zeta_{7} \, H_{4} + 2176 \, \zeta_{3}^2 \, H_{5} + 
 20472 \, \zeta_{6} \, H_{5} + 14688 \, \zeta_{2} \, \zeta_{3} \, H_{6} + 
 8592 \, \zeta_{5} \, H_{6} \nonumber \\ &
 + 55776 \, \zeta_{4} \, H_{7} + 31936 \, \zeta_{3} \, H_{8} + 52928 \, \zeta_{2} \, H_{9} + 
 64 \, \zeta_{3}^3 \, H_{0, 0} + 1896 \, \zeta_{4} \, \zeta_{5} \, H_{0, 0} \nonumber \\ &
 + 3792 \, \zeta_{3} \, \zeta_{6} \, H_{0, 0} + 576 \, \zeta_{2} \, \zeta_{7} \, 
 H_{0, 0} + 332 \, \zeta_{9} \, H_{0, 0} + 640 \, \zeta_{2} \, \zeta_{3}^2 \, H_{2, 0}
 + 800 \, \zeta_{3} \, \zeta_{5} \, H_{2, 0} \nonumber \\ &
 + \frac{6200}{3} \, \zeta_{8} \, H_{2, 0} + 
 3840 \, \zeta_{3} \, \zeta_{4} \, H_{2, 2} + 1920 \, \zeta_{2} \, \zeta_{5} \, 
  H_{2, 2} + 2880 \, \zeta_{7} \, H_{2, 2} + 
 960 \, \zeta_{3}^2 \, H_{2, 3}  \nonumber \\ &
 + 6120 \, \zeta_{6} \, H_{2, 3} + 5760 \, \zeta_{2} \, \zeta_{3} \, H_{2, 4}
 + 5280 \, \zeta_{5} \, H_{2, 4} + 17760 \, \zeta_{4} \, H_{2, 5}
 + 12160 \, \zeta_{3} \, H_{2, 6} \nonumber \\ &
  + 15680 \, \zeta_{2} \, H_{2, 7} + 7296 \, \zeta_{3} \, \zeta_{4} \, H_{3, 0} + 
 2304 \, \zeta_{2} \, \zeta_{5} \, H_{3, 0} + 1344 \, \zeta_{7} \, H_{3, 0} + 
 6144 \, \zeta_{3} \, \zeta_{4} \, H_{3, 1} \nonumber \\ &
 + 3072 \, \zeta_{2} \, \zeta_{5} \, 
  H_{3, 1} + 4608 \, \zeta_{7} \, H_{3, 1} + 
 1536 \, \zeta_{3}^2 \, H_{3, 2} + 9792 \, \zeta_{6} \, H_{3, 2} + 
 9216 \, \zeta_{2} \, \zeta_{3} \, H_{3, 3}  \nonumber \\ &
 + 8448 \, \zeta_{5} \, H_{3, 3} + 
 28416 \, \zeta_{4} \, H_{3, 4} + 19456 \, \zeta_{3} \, H_{3, 5} + 
 25088 \, \zeta_{2} \, H_{3, 6} + 1280 \, \zeta_{3}^2 \, H_{4, 0}  \nonumber \\ & + 
 18696 \, \zeta_{6} \, H_{4, 0} + 1728 \, \zeta_{3}^2 \, H_{4, 1} + 
 11016 \, \zeta_{6} \, H_{4, 1} + 11904 \, \zeta_{2} \, \zeta_{3} \, H_{4, 2} + 
 11808 \, \zeta_{5} \, H_{4, 2}  \nonumber \\ &
 + 35040 \, \zeta_{4} \, H_{4, 3} + 
 25664 \, \zeta_{3} \, H_{4, 4} + 31744 \, \zeta_{2} \, H_{4, 5} + 
 10944 \, \zeta_{2} \, \zeta_{3} \, H_{5, 0} + 5376 \, \zeta_{5} \, H_{5, 0} \nonumber \\ & + 
 14208 \, \zeta_{2} \, \zeta_{3} \, H_{5, 1} + 15936 \, \zeta_{5} \, H_{5, 1} + 
 38400 \, \zeta_{4} \, H_{5, 2} + 32064 \, \zeta_{3} \, H_{5, 3} + 
 36864 \, \zeta_{2} \, H_{5, 4} \nonumber \\ &
  + 56232 \, \zeta_{4} \, H_{6, 0} + 
 36960 \, \zeta_{4} \, H_{6, 1} + 38976 \, \zeta_{3} \, H_{6, 2} + 
 40128 \, \zeta_{2} \, H_{6, 3} + 22400 \, \zeta_{3} \, H_{7, 0}  \nonumber \\ & + 
 45248 \, \zeta_{3} \, H_{7, 1} + 39424 \, \zeta_{2} \, H_{7, 2} + 
 67712 \, \zeta_{2} \, H_{8, 0} + 29568 \, \zeta_{2} \, H_{8, 1} + 
 84752 \, H_{10, 0}  \nonumber \\ &
 + 736 \, \zeta_{2} \, \zeta_{3}^2 \, H_{0, 0, 0}
 + 608 \, \zeta_{3} \, \zeta_{5} \, H_{0, 0, 0} + 
 \frac{20012}{3} \, \zeta_{8} \, H_{0, 0, 0} + 4560 \, \zeta_{3} \, \zeta_{4} \, 
 H_{2, 0, 0} \nonumber \\ & 
 + 1440 \, \zeta_{2} \, \zeta_{5} \, H_{2, 0, 0} + 
 840 \, \zeta_{7} \, H_{2, 0, 0} + 640 \, \zeta_{3}^2 \, H_{2, 2, 0} + 
 4080 \, \zeta_{6} \, H_{2, 2, 0} + 3840 \, \zeta_{2} \, \zeta_{3} \, 
 H_{2, 2, 2} \nonumber \\ & + 5760 \, \zeta_{5} \, H_{2, 2, 2}
 + 7680 \, \zeta_{4} \, H_{2, 2, 3} + 7040 \, \zeta_{3} \, H_{2, 2, 4} + 
 6400 \, \zeta_{2} \, H_{2, 2, 5} + 3840 \, \zeta_{2} \, \zeta_{3} \, 
 H_{2, 3, 0} \nonumber \\ &
 + 2400 \, \zeta_{5} \, H_{2, 3, 0} + 5760 \, \zeta_{2} \, \zeta_{3} \, H_{2, 3, 1} + 
 8640 \, \zeta_{5} \, H_{2, 3, 1} + 11520 \, \zeta_{4} \, H_{2, 3, 2} \nonumber \\ & + 
 10560 \, \zeta_{3} \, H_{2, 3, 3} + 9600 \, \zeta_{2} \, H_{2, 3, 4} + 
 16560 \, \zeta_{4} \, H_{2, 4, 0} + 11520 \, \zeta_{4} \, H_{2, 4, 1} \nonumber \\ & + 
 14400 \, \zeta_{3} \, H_{2, 4, 2} + 11520 \, \zeta_{2} \, H_{2, 4, 3} + 
 7040 \, \zeta_{3} \, H_{2, 5, 0} + 18560 \, \zeta_{3} \, H_{2, 5, 1}  \nonumber \\ &+ 
 12160 \, \zeta_{2} \, H_{2, 5, 2} + 18080 \, \zeta_{2} \, H_{2, 6, 0} + 
 9600 \, \zeta_{2} \, H_{2, 6, 1} + 21440 \, H_{2, 8, 0} \nonumber \\ & + 
 768 \, \zeta_{3}^2 \, H_{3, 0, 0} + 15168 \, \zeta_{6} \, H_{3, 0, 0} + 
 1024 \, \zeta_{3}^2 \, H_{3, 1, 0} + 6528 \, \zeta_{6} \, H_{3, 1, 0} \nonumber \\ & + 
 6144 \, \zeta_{2} \, \zeta_{3} \, H_{3, 1, 2} + 
 9216 \, \zeta_{5} \, H_{3, 1, 2} + 12288 \, \zeta_{4} \, H_{3, 1, 3} + 
 11264 \, \zeta_{3} \, H_{3, 1, 4} \nonumber \\ & + 10240 \, \zeta_{2} \, H_{3, 1, 5} + 
 6144 \, \zeta_{2} \, \zeta_{3} \, H_{3, 2, 0} + 
 3840 \, \zeta_{5} \, H_{3, 2, 0} + 9216 \, \zeta_{2} \, \zeta_{3} \, 
  H_{3, 2, 1} \nonumber \\ & + 13824 \, \zeta_{5} \, H_{3, 2, 1}
  + 18432 \, \zeta_{4} \, H_{3, 2, 2} + 16896 \, \zeta_{3} \, H_{3, 2, 3} + 
 15360 \, \zeta_{2} \, H_{3, 2, 4} \nonumber \\ & + 26496 \, \zeta_{4} \, H_{3, 3, 0} + 
 18432 \, \zeta_{4} \, H_{3, 3, 1} + 23040 \, \zeta_{3} \, H_{3, 3, 2} + 
 18432 \, \zeta_{2} \, H_{3, 3, 3} \nonumber \\ & + 11264 \, \zeta_{3} \, H_{3, 4, 0} + 
 29696 \, \zeta_{3} \, H_{3, 4, 1} + 19456 \, \zeta_{2} \, H_{3, 4, 2} + 
 28928 \, \zeta_{2} \, H_{3, 5, 0} \nonumber \\ & + 15360 \, \zeta_{2} \, H_{3, 5, 1} + 
 34304 \, H_{3, 7, 0} + 8128 \, \zeta_{2} \, \zeta_{3} \, H_{4, 0, 0} + 
 3616 \, \zeta_{5} \, H_{4, 0, 0} \nonumber \\ & + 6912 \, \zeta_{2} \, \zeta_{3} \, 
  H_{4, 1, 0} + 4320 \, \zeta_{5} \, H_{4, 1, 0} + 
 10368 \, \zeta_{2} \, \zeta_{3} \, H_{4, 1, 1} + 
 15552 \, \zeta_{5} \, H_{4, 1, 1} \nonumber \\ & + 20736 \, \zeta_{4} \, H_{4, 1, 2} + 
 19008 \, \zeta_{3} \, H_{4, 1, 3} + 17280 \, \zeta_{2} \, H_{4, 1, 4} + 
 32112 \, \zeta_{4} \, H_{4, 2, 0} \nonumber \\ & + 20736 \, \zeta_{4} \, H_{4, 2, 1} + 
 28992 \, \zeta_{3} \, H_{4, 2, 2} + 22272 \, \zeta_{2} \, H_{4, 2, 3} + 
 14656 \, \zeta_{3} \, H_{4, 3, 0} \nonumber \\ & + 37504 \, \zeta_{3} \, H_{4, 3, 1} + 
 23936 \, \zeta_{2} \, H_{4, 3, 2} + 36544 \, \zeta_{2} \, H_{4, 4, 0} + 
 18816 \, \zeta_{2} \, H_{4, 4, 1} \nonumber \\ & + 43936 \, H_{4, 6, 0} + 
 51360 \, \zeta_{4} \, H_{5, 0, 0} + 33984 \, \zeta_{4} \, H_{5, 1, 0} + 
 18432 \, \zeta_{4} \, H_{5, 1, 1} \nonumber \\ & + 33024 \, \zeta_{3} \, H_{5, 1, 2} + 
 23424 \, \zeta_{2} \, H_{5, 1, 3} + 18048 \, \zeta_{3} \, H_{5, 2, 0} + 
 43008 \, \zeta_{3} \, H_{5, 2, 1} \nonumber \\ & + 26112 \, \zeta_{2} \, H_{5, 2, 2} + 
 42432 \, \zeta_{2} \, H_{5, 3, 0} + 20352 \, \zeta_{2} \, H_{5, 3, 1} + 
 52512 \, H_{5, 5, 0} \nonumber \\ & + 15776 \, \zeta_{3} \, H_{6, 0, 0} + 
 21120 \, \zeta_{3} \, H_{6, 1, 0} + 44160 \, \zeta_{3} \, H_{6, 1, 1} + 
 24960 \, \zeta_{2} \, H_{6, 1, 2} \nonumber \\ & + 46368 \, \zeta_{2} \, H_{6, 2, 0} + 
 19200 \, \zeta_{2} \, H_{6, 2, 1} + 60576 \, H_{6, 4, 0} + 
 66976 \, \zeta_{2} \, H_{7, 0, 0} \nonumber \\ & + 46144 \, \zeta_{2} \, H_{7, 1, 0} + 
 13440 \, \zeta_{2} \, H_{7, 1, 1} + 66976 \, H_{7, 3, 0}
 + 67712 \, H_{8, 2, 0} \nonumber \\ & + 105856 \, H_{9, 0, 0} + 
 52928 \, H_{9, 1, 0} + 8688 \, \zeta_{3} \, \zeta_{4} \, H_{0, 0, 0, 0} + 
 2064 \, \zeta_{2} \, \zeta_{5} \, H_{0, 0, 0, 0} \nonumber \\ & + 
 792 \, \zeta_{7} \, H_{0, 0, 0, 0} + 480 \, \zeta_{3}^2 \, H_{2, 0, 0, 0} + 
 9480 \, \zeta_{6} \, H_{2, 0, 0, 0} + 2560 \, \zeta_{2} \, \zeta_{3} \, 
  H_{2, 2, 0, 0} \nonumber \\ & + 1600 \, \zeta_{5} \, H_{2, 2, 0, 0} + 
 5760 \, \zeta_{4} \, H_{2, 2, 2, 0} + 7680 \, \zeta_{3} \, H_{2, 2, 2, 2} + 
 3840 \, \zeta_{2} \, H_{2, 2, 2, 3} \nonumber \\ & + 2560 \, \zeta_{3} \, H_{2, 2, 3, 0} + 
 10240 \, \zeta_{3} \, H_{2, 2, 3, 1} + 5120 \, \zeta_{2} \, H_{2, 2, 3, 2} + 
 6400 \, \zeta_{2} \, H_{2, 2, 4, 0} \nonumber \\ & + 3840 \, \zeta_{2} \, H_{2, 2, 4, 1} + 
 7360 \, H_{2, 2, 6, 0} + 13680 \, \zeta_{4} \, H_{2, 3, 0, 0} + 
 8640 \, \zeta_{4} \, H_{2, 3, 1, 0} \nonumber \\ & + 11520 \, \zeta_{3} \, H_{2, 3, 1, 2} + 
 5760 \, \zeta_{2} \, H_{2, 3, 1, 3} + 3840 \, \zeta_{3} \, H_{2, 3, 2, 0} + 
 15360 \, \zeta_{3} \, H_{2, 3, 2, 1} \nonumber \\ & + 7680 \, \zeta_{2} \, H_{2, 3, 2, 2} + 
 9600 \, \zeta_{2} \, H_{2, 3, 3, 0} + 5760 \, \zeta_{2} \, H_{2, 3, 3, 1} + 
 11040 \, H_{2, 3, 5, 0} \nonumber \\ & + 4480 \, \zeta_{3} \, H_{2, 4, 0, 0} + 
 3840 \, \zeta_{3} \, H_{2, 4, 1, 0} + 15360 \, \zeta_{3} \, H_{2, 4, 1, 1} + 
 7680 \, \zeta_{2} \, H_{2, 4, 1, 2} \nonumber \\ & + 11520 \, \zeta_{2} \, H_{2, 4, 2, 0} + 
 5760 \, \zeta_{2} \, H_{2, 4, 2, 1} + 13920 \, H_{2, 4, 4, 0} + 
 16480 \, \zeta_{2} \, H_{2, 5, 0, 0} \nonumber \\ & + 
 12160 \, \zeta_{2} \, H_{2, 5, 1, 0} + 3840 \, \zeta_{2} \, H_{2, 5, 1, 1} + 
 16480 \, H_{2, 5, 3, 0} + 18080 \, H_{2, 6, 2, 0} \nonumber \\ & + 
 23520 \, H_{2, 7, 0, 0} + 15680 \, H_{2, 7, 1, 0} + 
 5888 \, \zeta_{2} \, \zeta_{3} \, H_{3, 0, 0, 0} + 
 2432 \, \zeta_{5} \, H_{3, 0, 0, 0} \nonumber \\ & + 4096 \, \zeta_{2} \, \zeta_{3} \, 
  H_{3, 1, 0, 0} + 2560 \, \zeta_{5} \, H_{3, 1, 0, 0} + 
 9216 \, \zeta_{4} \, H_{3, 1, 2, 0} + 12288 \, \zeta_{3} \, H_{3, 1, 2, 2} \nonumber \\ & + 
 6144 \, \zeta_{2} \, H_{3, 1, 2, 3} + 4096 \, \zeta_{3} \, H_{3, 1, 3, 0} + 
 16384 \, \zeta_{3} \, H_{3, 1, 3, 1} + 8192 \, \zeta_{2} \, H_{3, 1, 3, 2} \nonumber \\ & + 
 10240 \, \zeta_{2} \, H_{3, 1, 4, 0} + 6144 \, \zeta_{2} \, H_{3, 1, 4, 1} + 
 11776 \, H_{3, 1, 6, 0} + 21888 \, \zeta_{4} \, H_{3, 2, 0, 0} \nonumber \\ & + 
 13824 \, \zeta_{4} \, H_{3, 2, 1, 0} + 
 18432 \, \zeta_{3} \, H_{3, 2, 1, 2} + 9216 \, \zeta_{2} \, H_{3, 2, 1, 3} + 
 6144 \, \zeta_{3} \, H_{3, 2, 2, 0} \nonumber \\ & + 24576 \, \zeta_{3} \, H_{3, 2, 2, 1} + 
 12288 \, \zeta_{2} \, H_{3, 2, 2, 2} + 
 15360 \, \zeta_{2} \, H_{3, 2, 3, 0} + 9216 \, \zeta_{2} \, H_{3, 2, 3, 1} \nonumber \\ & + 
 17664 \, H_{3, 2, 5, 0} + 7168 \, \zeta_{3} \, H_{3, 3, 0, 0} + 
 6144 \, \zeta_{3} \, H_{3, 3, 1, 0} + 24576 \, \zeta_{3} \, H_{3, 3, 1, 1} \nonumber \\ & + 
 12288 \, \zeta_{2} \, H_{3, 3, 1, 2} + 
 18432 \, \zeta_{2} \, H_{3, 3, 2, 0} + 9216 \, \zeta_{2} \, H_{3, 3, 2, 1} + 
 22272 \, H_{3, 3, 4, 0} \nonumber \\ & + 26368 \, \zeta_{2} \, H_{3, 4, 0, 0} + 
 19456 \, \zeta_{2} \, H_{3, 4, 1, 0} + 6144 \, \zeta_{2} \, H_{3, 4, 1, 1} + 
 26368 \, H_{3, 4, 3, 0} \nonumber \\ & + 28928 \, H_{3, 5, 2, 0} + 
 37632 \, H_{3, 6, 0, 0} + 25088 \, H_{3, 6, 1, 0} + 
 44064 \, \zeta_{4} \, H_{4, 0, 0, 0} \nonumber \\ & + 
 24624 \, \zeta_{4} \, H_{4, 1, 0, 0} + 
 15552 \, \zeta_{4} \, H_{4, 1, 1, 0} + 
 20736 \, \zeta_{3} \, H_{4, 1, 1, 2} + 
 10368 \, \zeta_{2} \, H_{4, 1, 1, 3} \nonumber \\ & +
 6912 \, \zeta_{3} \, H_{4, 1, 2, 0} + 
 27648 \, \zeta_{3} \, H_{4, 1, 2, 1} + 
 13824 \, \zeta_{2} \, H_{4, 1, 2, 2} + 
 17280 \, \zeta_{2} \, H_{4, 1, 3, 0} \nonumber \\ &
 + 10368 \, \zeta_{2} \, H_{4, 1, 3, 1} + 19872 \, H_{4, 1, 5, 0} + 
 9472 \, \zeta_{3} \, H_{4, 2, 0, 0} + 6912 \, \zeta_{3} \, H_{4, 2, 1, 0} \nonumber \\ & + 
 27648 \, \zeta_{3} \, H_{4, 2, 1, 1} + 
 13824 \, \zeta_{2} \, H_{4, 2, 1, 2} + 
 22272 \, \zeta_{2} \, H_{4, 2, 2, 0} + 
 10368 \, \zeta_{2} \, H_{4, 2, 2, 1} \nonumber \\ & + 27552 \, H_{4, 2, 4, 0} + 
 33152 \, \zeta_{2} \, H_{4, 3, 0, 0} + 
 23936 \, \zeta_{2} \, H_{4, 3, 1, 0} + 6912 \, \zeta_{2} \, H_{4, 3, 1, 1} \nonumber \\ & + 
 33152 \, H_{4, 3, 3, 0} + 36544 \, H_{4, 4, 2, 0} + 
 47936 \, H_{4, 5, 0, 0} + 31744 \, H_{4, 5, 1, 0} \nonumber \\ & + 
 11264 \, \zeta_{3} \, H_{5, 0, 0, 0} + 
 11968 \, \zeta_{3} \, H_{5, 1, 0, 0} + 6144 \, \zeta_{3} \, H_{5, 1, 1, 0} + 
 24576 \, \zeta_{3} \, H_{5, 1, 1, 1} \nonumber \\ & + 
 12288 \, \zeta_{2} \, H_{5, 1, 1, 2} + 
 23424 \, \zeta_{2} \, H_{5, 1, 2, 0} + 9216 \, \zeta_{2} \, H_{5, 1, 2, 1} + 
 30720 \, H_{5, 1, 4, 0} \nonumber \\ & + 38208 \, \zeta_{2} \, H_{5, 2, 0, 0} + 
 26112 \, \zeta_{2} \, H_{5, 2, 1, 0} + 6144 \, \zeta_{2} \, H_{5, 2, 1, 1} + 
 38208 \, H_{5, 2, 3, 0} \nonumber \\ & + 42432 \, H_{5, 3, 2, 0} + 
 56896 \, H_{5, 4, 0, 0} + 36864 \, H_{5, 4, 1, 0} + 
 60576 \, \zeta_{2} \, H_{6, 0, 0, 0} \nonumber \\ & + 
 40800 \, \zeta_{2} \, H_{6, 1, 0, 0} + 
 24960 \, \zeta_{2} \, H_{6, 1, 1, 0} + 3840 \, \zeta_{2} \, H_{6, 1, 1, 1} + 
 40800 \, H_{6, 1, 3, 0} \nonumber \\ & + 46368 \, H_{6, 2, 2, 0} + 
 65248 \, H_{6, 3, 0, 0} + 40128 \, H_{6, 3, 1, 0} + 
 46144 \, H_{7, 1, 2, 0} \nonumber \\ & + 72128 \, H_{7, 2, 0, 0} + 
 39424 \, H_{7, 2, 1, 0} + 102640 \, H_{8, 0, 0, 0} + 
 73920 \, H_{8, 1, 0, 0} \nonumber \\ & + 29568 \, H_{8, 1, 1, 0} + 
 736 \, \zeta_{3}^2 \, H_{0, 0, 0, 0, 0} + 
 27736 \, \zeta_{6} \, H_{0, 0, 0, 0, 0} + 3680 \, \zeta_{2} \, \zeta_{3} \, 
  H_{2, 0, 0, 0, 0} \nonumber \\ & + 1520 \, \zeta_{5} \, H_{2, 0, 0, 0, 0} + 
 9120 \, \zeta_{4} \, H_{2, 2, 0, 0, 0} + 
 1920 \, \zeta_{3} \, H_{2, 2, 2, 0, 0} + 
 3840 \, \zeta_{2} \, H_{2, 2, 2, 2, 0} \nonumber \\ & + 3840 \, H_{2, 2, 2, 4, 0} + 
 5120 \, \zeta_{2} \, H_{2, 2, 3, 0, 0} + 
 5120 \, \zeta_{2} \, H_{2, 2, 3, 1, 0} + 5120 \, H_{2, 2, 3, 3, 0} \nonumber \\ & + 
 6400 \, H_{2, 2, 4, 2, 0} + 6400 \, H_{2, 2, 5, 0, 0} + 
 6400 \, H_{2, 2, 5, 1, 0} + 2880 \, \zeta_{3} \, H_{2, 3, 0, 0, 0} \nonumber \\ & + 
 2880 \, \zeta_{3} \, H_{2, 3, 1, 0, 0} + 
 5760 \, \zeta_{2} \, H_{2, 3, 1, 2, 0} + 5760 \, H_{2, 3, 1, 4, 0} + 
 7680 \, \zeta_{2} \, H_{2, 3, 2, 0, 0} \nonumber \\ &
 + 7680 \, \zeta_{2} \, H_{2, 3, 2, 1, 0} + 7680 \, H_{2, 3, 2, 3, 0} + 
 9600 \, H_{2, 3, 3, 2, 0} + 9600 \, H_{2, 3, 4, 0, 0} \nonumber \\ & + 
 9600 \, H_{2, 3, 4, 1, 0} + 13920 \, \zeta_{2} \, H_{2, 4, 0, 0, 0} + 
 7680 \, \zeta_{2} \, H_{2, 4, 1, 0, 0} + 
 7680 \, \zeta_{2} \, H_{2, 4, 1, 1, 0} \nonumber \\ & + 7680 \, H_{2, 4, 1, 3, 0} + 
 11520 \, H_{2, 4, 2, 2, 0} + 11840 \, H_{2, 4, 3, 0, 0} + 
 11520 \, H_{2, 4, 3, 1, 0} \nonumber \\ & + 12160 \, H_{2, 5, 1, 2, 0} + 
 13760 \, H_{2, 5, 2, 0, 0} + 12160 \, H_{2, 5, 2, 1, 0} + 
 20800 \, H_{2, 6, 0, 0, 0} \nonumber \\ & + 14400 \, H_{2, 6, 1, 0, 0} + 
 9600 \, H_{2, 6, 1, 1, 0} + 34752 \, \zeta_{4} \, H_{3, 0, 0, 0, 0} + 
 14592 \, \zeta_{4} \, H_{3, 1, 0, 0, 0} \nonumber \\ & + 
 3072 \, \zeta_{3} \, H_{3, 1, 2, 0, 0} + 
 6144 \, \zeta_{2} \, H_{3, 1, 2, 2, 0} + 6144 \, H_{3, 1, 2, 4, 0} + 
 8192 \, \zeta_{2} \, H_{3, 1, 3, 0, 0} \nonumber \\ & + 
 8192 \, \zeta_{2} \, H_{3, 1, 3, 1, 0} + 8192 \, H_{3, 1, 3, 3, 0} + 
 10240 \, H_{3, 1, 4, 2, 0} + 10240 \, H_{3, 1, 5, 0, 0} \nonumber \\ & + 
 10240 \, H_{3, 1, 5, 1, 0} + 4608 \, \zeta_{3} \, H_{3, 2, 0, 0, 0} + 
 4608 \, \zeta_{3} \, H_{3, 2, 1, 0, 0} + 
 9216 \, \zeta_{2} \, H_{3, 2, 1, 2, 0} \nonumber \\ & + 9216 \, H_{3, 2, 1, 4, 0} + 
 12288 \, \zeta_{2} \, H_{3, 2, 2, 0, 0} + 
 12288 \, \zeta_{2} \, H_{3, 2, 2, 1, 0} + 12288 \, H_{3, 2, 2, 3, 0} \nonumber \\ & + 
 15360 \, H_{3, 2, 3, 2, 0} + 15360 \, H_{3, 2, 4, 0, 0} + 
 15360 \, H_{3, 2, 4, 1, 0} + 22272 \, \zeta_{2} \, H_{3, 3, 0, 0, 0} \nonumber \\ & + 
 12288 \, \zeta_{2} \, H_{3, 3, 1, 0, 0} + 
 12288 \, \zeta_{2} \, H_{3, 3, 1, 1, 0} + 12288 \, H_{3, 3, 1, 3, 0} + 
 18432 \, H_{3, 3, 2, 2, 0} \nonumber \\ & + 18944 \, H_{3, 3, 3, 0, 0} + 
 18432 \, H_{3, 3, 3, 1, 0} + 19456 \, H_{3, 4, 1, 2, 0} + 
 22016 \, H_{3, 4, 2, 0, 0} \nonumber \\ & + 19456 \, H_{3, 4, 2, 1, 0} + 
 33280 \, H_{3, 5, 0, 0, 0} + 23040 \, H_{3, 5, 1, 0, 0} + 
 15360 \, H_{3, 5, 1, 1, 0} \nonumber \\ & + 8192 \, \zeta_{3} \, H_{4, 0, 0, 0, 0} + 
 5184 \, \zeta_{3} \, H_{4, 1, 0, 0, 0} + 
 5184 \, \zeta_{3} \, H_{4, 1, 1, 0, 0} + 
 10368 \, \zeta_{2} \, H_{4, 1, 1, 2, 0} \nonumber \\ & + 10368 \, H_{4, 1, 1, 4, 0} + 
 13824 \, \zeta_{2} \, H_{4, 1, 2, 0, 0} + 
 13824 \, \zeta_{2} \, H_{4, 1, 2, 1, 0} + 13824 \, H_{4, 1, 2, 3, 0} \nonumber \\ & + 
 17280 \, H_{4, 1, 3, 2, 0} + 17280 \, H_{4, 1, 4, 0, 0} + 
 17280 \, H_{4, 1, 4, 1, 0} + 27552 \, \zeta_{2} \, H_{4, 2, 0, 0, 0} \nonumber \\ & + 
 13824 \, \zeta_{2} \, H_{4, 2, 1, 0, 0} + 
 13824 \, \zeta_{2} \, H_{4, 2, 1, 1, 0} + 13824 \, H_{4, 2, 1, 3, 0} + 
 22272 \, H_{4, 2, 2, 2, 0} \nonumber \\ & + 23360 \, H_{4, 2, 3, 0, 0} + 
 22272 \, H_{4, 2, 3, 1, 0} + 23936 \, H_{4, 3, 1, 2, 0} + 
 27712 \, H_{4, 3, 2, 0, 0} \nonumber \\ & + 23936 \, H_{4, 3, 2, 1, 0} + 
 42368 \, H_{4, 4, 0, 0, 0} + 28608 \, H_{4, 4, 1, 0, 0} + 
 18816 \, H_{4, 4, 1, 1, 0} \nonumber \\ & + 52512 \, \zeta_{2} \, H_{5, 0, 0, 0, 0} + 
 30720 \, \zeta_{2} \, H_{5, 1, 0, 0, 0} + 
 12288 \, \zeta_{2} \, H_{5, 1, 1, 0, 0} + 
 12288 \, \zeta_{2} \, H_{5, 1, 1, 1, 0} \nonumber \\ & + 12288 \, H_{5, 1, 1, 3, 0}
 + 23424 \, H_{5, 1, 2, 2, 0} + 26048 \, H_{5, 1, 3, 0, 0} + 
 23424 \, H_{5, 1, 3, 1, 0} \nonumber \\ & + 26112 \, H_{5, 2, 1, 2, 0} + 
 32256 \, H_{5, 2, 2, 0, 0} + 26112 \, H_{5, 2, 2, 1, 0} + 
 50304 \, H_{5, 3, 0, 0, 0} \nonumber \\ & + 32448 \, H_{5, 3, 1, 0, 0} + 
 20352 \, H_{5, 3, 1, 1, 0} + 24960 \, H_{6, 1, 1, 2, 0} + 
 35520 \, H_{6, 1, 2, 0, 0} \nonumber \\ & + 24960 \, H_{6, 1, 2, 1, 0} + 
 57792 \, H_{6, 2, 0, 0, 0} + 34560 \, H_{6, 2, 1, 0, 0} + 
 19200 \, H_{6, 2, 1, 1, 0} \nonumber \\ & + 91392 \, H_{7, 0, 0, 0, 0} + 
 64064 \, H_{7, 1, 0, 0, 0} + 33600 \, H_{7, 1, 1, 0, 0} + 
 13440 \, H_{7, 1, 1, 1, 0} \nonumber \\ & + 8576 \, \zeta_{2} \, \zeta_{3} \, 
  H_{0, 0, 0, 0, 0, 0} + 2672 \, \zeta_{5} \, H_{0, 0, 0, 0, 0, 0} + 
 21720 \, \zeta_{4} \, H_{2, 0, 0, 0, 0, 0} \nonumber \\ & + 
 3840 \, \zeta_{2} \, H_{2, 2, 2, 0, 0, 0} + 
 3840 \, H_{2, 2, 2, 2, 2, 0} + 1920 \, H_{2, 2, 2, 3, 0, 0} + 
 3840 \, H_{2, 2, 2, 3, 1, 0} \nonumber \\ & + 5120 \, H_{2, 2, 3, 1, 2, 0} + 
 2560 \, H_{2, 2, 3, 2, 0, 0} + 5120 \, H_{2, 2, 3, 2, 1, 0} + 
 5120 \, H_{2, 2, 4, 0, 0, 0} \nonumber \\ & + 1920 \, H_{2, 2, 4, 1, 0, 0} + 
 3840 \, H_{2, 2, 4, 1, 1, 0} + 
 11040 \, \zeta_{2} \, H_{2, 3, 0, 0, 0, 0} + 
 5760 \, \zeta_{2} \, H_{2, 3, 1, 0, 0, 0} \nonumber \\ & + 
 1920 \, \zeta_{3} \, H_{2, 2, 0, 0, 0, 0} + 
 5760 \, H_{2, 3, 1, 2, 2, 0} + 2880 \, H_{2, 3, 1, 3, 0, 0} + 
 5760 \, H_{2, 3, 1, 3, 1, 0} \nonumber \\ & + 7680 \, H_{2, 3, 2, 1, 2, 0} + 
 3840 \, H_{2, 3, 2, 2, 0, 0} + 7680 \, H_{2, 3, 2, 2, 1, 0} + 
 7680 \, H_{2, 3, 3, 0, 0, 0} \nonumber \\ & + 2880 \, H_{2, 3, 3, 1, 0, 0} + 
 5760 \, H_{2, 3, 3, 1, 1, 0} + 7680 \, H_{2, 4, 1, 1, 2, 0} + 
 3840 \, H_{2, 4, 1, 2, 0, 0} \nonumber \\ & + 7680 \, H_{2, 4, 1, 2, 1, 0} + 
 9600 \, H_{2, 4, 2, 0, 0, 0} + 2880 \, H_{2, 4, 2, 1, 0, 0} + 
 5760 \, H_{2, 4, 2, 1, 1, 0} \nonumber \\ & + 17280 \, H_{2, 5, 0, 0, 0, 0} + 
 11360 \, H_{2, 5, 1, 0, 0, 0} + 1920 \, H_{2, 5, 1, 1, 0, 0} + 
 3840 \, H_{2, 5, 1, 1, 1, 0} \nonumber \\ & + 
 5888 \, \zeta_{3} \, H_{3, 0, 0, 0, 0, 0} + 
 3072 \, \zeta_{3} \, H_{3, 1, 0, 0, 0, 0} + 
 6144 \, \zeta_{2} \, H_{3, 1, 2, 0, 0, 0} + 
 6144 \, H_{3, 1, 2, 2, 2, 0} \nonumber \\ & + 3072 \, H_{3, 1, 2, 3, 0, 0} + 
 6144 \, H_{3, 1, 2, 3, 1, 0} + 8192 \, H_{3, 1, 3, 1, 2, 0} + 
 4096 \, H_{3, 1, 3, 2, 0, 0} \nonumber \\ & + 8192 \, H_{3, 1, 3, 2, 1, 0} + 
 8192 \, H_{3, 1, 4, 0, 0, 0} + 3072 \, H_{3, 1, 4, 1, 0, 0} + 
 6144 \, H_{3, 1, 4, 1, 1, 0} \nonumber \\ & + 
 17664 \, \zeta_{2} \, H_{3, 2, 0, 0, 0, 0} + 
 9216 \, \zeta_{2} \, H_{3, 2, 1, 0, 0, 0} + 
 9216 \, H_{3, 2, 1, 2, 2, 0} + 4608 \, H_{3, 2, 1, 3, 0, 0} \nonumber \\ & + 
 9216 \, H_{3, 2, 1, 3, 1, 0} + 12288 \, H_{3, 2, 2, 1, 2, 0} + 
 6144 \, H_{3, 2, 2, 2, 0, 0} + 12288 \, H_{3, 2, 2, 2, 1, 0} \nonumber \\ & + 
 12288 \, H_{3, 2, 3, 0, 0, 0} + 4608 \, H_{3, 2, 3, 1, 0, 0} + 
 9216 \, H_{3, 2, 3, 1, 1, 0} + 12288 \, H_{3, 3, 1, 1, 2, 0} \nonumber \\ & + 
 6144 \, H_{3, 3, 1, 2, 0, 0} + 12288 \, H_{3, 3, 1, 2, 1, 0} + 
 15360 \, H_{3, 3, 2, 0, 0, 0} + 4608 \, H_{3, 3, 2, 1, 0, 0} \nonumber \\ & + 
 9216 \, H_{3, 3, 2, 1, 1, 0} + 27648 \, H_{3, 4, 0, 0, 0, 0} + 
 18176 \, H_{3, 4, 1, 0, 0, 0} + 3072 \, H_{3, 4, 1, 1, 0, 0} \nonumber \\ & + 
 6144 \, H_{3, 4, 1, 1, 1, 0} + 
 43936 \, \zeta_{2} \, H_{4, 0, 0, 0, 0, 0} + 
 19872 \, \zeta_{2} \, H_{4, 1, 0, 0, 0, 0} + 
 10368 \, \zeta_{2} \, H_{4, 1, 1, 0, 0, 0} \nonumber \\ & 
 + 10368 \, H_{4, 1, 1, 2, 2, 0} + 5184 \, H_{4, 1, 1, 3, 0, 0} + 
 10368 \, H_{4, 1, 1, 3, 1, 0} + 13824 \, H_{4, 1, 2, 1, 2, 0} \nonumber \\ & + 
 6912 \, H_{4, 1, 2, 2, 0, 0} + 13824 \, H_{4, 1, 2, 2, 1, 0} + 
 13824 \, H_{4, 1, 3, 0, 0, 0} + 5184 \, H_{4, 1, 3, 1, 0, 0} \nonumber \\ & + 
 10368 \, H_{4, 1, 3, 1, 1, 0} + 13824 \, H_{4, 2, 1, 1, 2, 0} + 
 6912 \, H_{4, 2, 1, 2, 0, 0} + 13824 \, H_{4, 2, 1, 2, 1, 0} \nonumber \\ & + 
 19008 \, H_{4, 2, 2, 0, 0, 0} + 5184 \, H_{4, 2, 2, 1, 0, 0} + 
 10368 \, H_{4, 2, 2, 1, 1, 0} + 35040 \, H_{4, 3, 0, 0, 0, 0} \nonumber \\ & + 
 22912 \, H_{4, 3, 1, 0, 0, 0} + 3456 \, H_{4, 3, 1, 1, 0, 0} + 
 6912 \, H_{4, 3, 1, 1, 1, 0} + 12288 \, H_{5, 1, 1, 1, 2, 0} \nonumber \\ & + 
 6144 \, H_{5, 1, 1, 2, 0, 0} + 12288 \, H_{5, 1, 1, 2, 1, 0} + 
 21312 \, H_{5, 1, 2, 0, 0, 0} + 4608 \, H_{5, 1, 2, 1, 0, 0} \nonumber \\ & + 
 9216 \, H_{5, 1, 2, 1, 1, 0} + 41280 \, H_{5, 2, 0, 0, 0, 0} + 
 26688 \, H_{5, 2, 1, 0, 0, 0} + 3072 \, H_{5, 2, 1, 1, 0, 0} \nonumber \\ & + 
 6144 \, H_{5, 2, 1, 1, 1, 0} + 78656 \, H_{6, 0, 0, 0, 0, 0} + 
 46080 \, H_{6, 1, 0, 0, 0, 0} + 29280 \, H_{6, 1, 1, 0, 0, 0} \nonumber \\ & + 
 1920 \, H_{6, 1, 1, 1, 0, 0} + 3840 \, H_{6, 1, 1, 1, 1, 0} + 
 72000 \, \zeta_{4} \, H_{0, 0, 0, 0, 0, 0, 0} + 
 3680 \, \zeta_{3} \, H_{2, 0, 0, 0, 0, 0, 0} \nonumber \\ & + 
 7360 \, \zeta_{2} \, H_{2, 2, 0, 0, 0, 0, 0} + 
 1920 \, H_{2, 2, 2, 2, 0, 0, 0} + 
 3840 \, H_{2, 2, 3, 0, 0, 0, 0} + 
 2560 \, H_{2, 2, 3, 1, 0, 0, 0} \nonumber \\ & + 
 2880 \, H_{2, 3, 1, 2, 0, 0, 0} + 
 5760 \, H_{2, 3, 2, 0, 0, 0, 0} + 
 3840 \, H_{2, 3, 2, 1, 0, 0, 0} + 
 14080 \, H_{2, 4, 0, 0, 0, 0, 0} \nonumber \\ & + 
 5760 \, H_{2, 4, 1, 0, 0, 0, 0} + 
 3840 \, H_{2, 4, 1, 1, 0, 0, 0} + 
 34304 \, \zeta_{2} \, H_{3, 0, 0, 0, 0, 0, 0} + 
 11776 \, \zeta_{2} \, H_{3, 1, 0, 0, 0, 0, 0} \nonumber \\ & + 
 3072 \, H_{3, 1, 2, 2, 0, 0, 0} + 
 6144 \, H_{3, 1, 3, 0, 0, 0, 0} + 
 4096 \, H_{3, 1, 3, 1, 0, 0, 0} + 
 4608 \, H_{3, 2, 1, 2, 0, 0, 0} \nonumber \\ & + 
 9216 \, H_{3, 2, 2, 0, 0, 0, 0} + 
 6144 \, H_{3, 2, 2, 1, 0, 0, 0} + 
 22528 \, H_{3, 3, 0, 0, 0, 0, 0} + 
 9216 \, H_{3, 3, 1, 0, 0, 0, 0} \nonumber \\ & + 
 6144 \, H_{3, 3, 1, 1, 0, 0, 0} + 
 5184 \, H_{4, 1, 1, 2, 0, 0, 0} + 
 10368 \, H_{4, 1, 2, 0, 0, 0, 0} + 
 6912 \, H_{4, 1, 2, 1, 0, 0, 0} \nonumber \\ & + 
 28096 \, H_{4, 2, 0, 0, 0, 0, 0} + 
 10368 \, H_{4, 2, 1, 0, 0, 0, 0} + 
 6912 \, H_{4, 2, 1, 1, 0, 0, 0} + 
 66560 \, H_{5, 0, 0, 0, 0, 0, 0} \nonumber \\ & + 
 32032 \, H_{5, 1, 0, 0, 0, 0, 0} + 
 9216 \, H_{5, 1, 1, 0, 0, 0, 0} + 
 6144 \, H_{5, 1, 1, 1, 0, 0, 0} \nonumber \\ & + 
 10720 \, \zeta_{3} \, H_{0, 0, 0, 0, 0, 0, 0, 0} + 
 21440 \, \zeta_{2} \, H_{2, 0, 0, 0, 0, 0, 0, 0} + 
 2880 \, H_{2, 2, 2, 0, 0, 0, 0, 0} \nonumber \\ & + 
 11040 \, H_{2, 3, 0, 0, 0, 0, 0, 0} + 
 4320 \, H_{2, 3, 1, 0, 0, 0, 0, 0} + 
 4608 \, H_{3, 1, 2, 0, 0, 0, 0, 0} \nonumber \\ & + 
 17664 \, H_{3, 2, 0, 0, 0, 0, 0, 0} + 
 6912 \, H_{3, 2, 1, 0, 0, 0, 0, 0} + 
 55104 \, H_{4, 0, 0, 0, 0, 0, 0, 0} \nonumber \\ & + 
 19872 \, H_{4, 1, 0, 0, 0, 0, 0, 0} + 
 7776 \, H_{4, 1, 1, 0, 0, 0, 0, 0} + 
 84752 \, \zeta_{2} \, H_{0, 0, 0, 0, 0, 0, 0, 0, 0} \nonumber \\ & + 
 7360 \, H_{2, 2, 0, 0, 0, 0, 0, 0, 0} + 
 42880 \, H_{3, 0, 0, 0, 0, 0, 0, 0, 0} + 
 11776 \, H_{3, 1, 0, 0, 0, 0, 0, 0, 0} \nonumber \\ & + 
 26800 \, H_{2, 0, 0, 0, 0, 0, 0, 0, 0, 0} + 
 127128 \, H_{0, 0, 0, 0, 0, 0, 0, 0, 0, 0, 0} \; .
\end{align}
}

\end{document}